\documentclass[structabstract]{aa}  
\usepackage{natbib}
\usepackage{graphicx}
\usepackage{multirow}
\usepackage{txfonts}
\begin{document}
   \title{$\Delta Y/ \Delta Z$ from the analysis of local K dwarfs}


\author{M. Gennaro \inst{1}
	\thanks{Member of the International Max Planck Research School for Astronomy and Cosmic Physics at the University of Heidelberg, IMPRS-HD, Germany}
           \and
           P. G. Prada Moroni \inst{2,3}
          \and
          S. Degl'Innocenti \inst{2,3}
                    }

\institute{Max Planck Institute for Astronomy, K\"onigsthul 17, D-69117 Heidelberg,
     Germany\\
\email{gennaro@mpia.de}
	\and
	Physics Department ``E. Fermi'', University of Pisa, Largo B. Pontecorvo 3, I-56127, Pisa, Italy
         \and
	INFN, Largo B. Pontecorvo 3, I-56127, Pisa, Italy\\
             \email{prada@df.unipi.it, scilla@df.unipi.it}
}

   \date{Received February 9, 2010; accepted April 29, 2010}

 
  \abstract
   {The stellar helium-to-metal enrichment ratio, $\Delta Y/ \Delta Z$, is a widely studied astrophysical quantity. However, its value is still not precisely constrained.} 
   {This paper is focused on the study of the main sources of  uncertainty which affect the $\Delta Y/ \Delta Z$ ratio derived from the analysis of the low-main sequence (MS) stars in the solar neighborhood.}
   {The possibility to infer the value of the helium-to-metal enrichment ratio from the study of low-MS stars relies on the dependence of the stellar luminosity and effective temperature on the initial helium and metal abundances. The $\Delta Y/\Delta Z$ ratio is obtained by comparing the magnitude difference between the observed stars and a reference theoretical zero age main sequence (ZAMS) with the related theoretical magnitude differences computed from a new set of stellar models with up-to-date input physics and a fine grid of chemical compositions. A Monte Carlo approach has been used to evaluate the impact on the result of different sources of uncertainty, i.e. observational errors, evolutionary effects, systematic uncertainties of the models. As a check of the procedure, the method has been applied to a different data set, namely the low-MS of the Hyades.}
   {Once a set of ZAMS and atmosphere models have been chosen, we found that the inferred value of $\Delta Y/ \Delta Z$ is sensitive to the age of the stellar sample, even if we restricted the data set to low luminosity stars. The lack of an accurate age estimate of low mass field stars leads to an underestimate of the inferred $\Delta Y/ \Delta Z$ of $\sim 2$ units. On the contrary the method firmly recovers the $\Delta Y/ \Delta Z$ value for not evolved samples of stars such as the Hyades low-MS.
  Adopting a solar calibrated mixing-length parameter and the PHOENIX GAIA v2.6.1 atmospheric models, we found $\Delta Y/\Delta Z~= 5.3~\pm~1.4$ once the age correction has been applied. The Hyades sample provided a perfectly consistent value.  
}
   {We have demonstrated that the assumption that low-mass stars in the solar
     neighborhood can be considered as unevolved, does not necessarily hold,
     and it may indeed lead to a bias in the inferred $\Delta Y/\Delta Z$.  The effect of the still poorly constrained efficiency of the superadiabatic convection and of different atmosphere models adopted to transform luminosities and effective temperature into colors and magnitudes have been discussed, too.}

   \keywords{Galaxy: fundamental parameters -- solar neighborhood --
                Stars: abundances --
                Stars: low-mass
               }

   \maketitle
%

\section{Introduction}
It is a well known and firm result of stellar evolution studies that the main structural, observational and evolutionary characteristics of 
a star of given mass depend sensitively on the original chemical composition, i.e. the initial helium and metal abundances, $Y$ and $Z$, respectively. 
As a consequence, these parameters affect also the observable quantities of stellar systems, from star clusters to galaxies.

While the present $Z$ in a stellar atmosphere can be obtained by the direct spectroscopic measurements of some tracer element, mainly iron, with the additional assumption on the mixture of heavy elements, the situation of $Y$ is completely different. In the vast majority of stars the helium lines can not be observed, with the exception of those hotter than $20\,000$ K. This means that for the low-mass stars, which are the most common and long-lasting objects in the Universe, helium can be directly observed only in advanced evolutionary phases, as the blue part of the horizontal branch or in the post-asymptotic giant branch.
Thus, the  actually measured helium abundance is not the original one, but the result of several complex processes, like dredge-up of nuclearly processed material, diffusion and radiative levitation, which severely alter the surface chemical composition. 

As a consequence, in order to evaluate the original $Y$, the only possibility is to rely on indirect methods. This explains why such an important parameter is still poorly constrained.

As early suggested by \citet{peimbert74} from the analysis of the chemical composition of the HII regions in the Large Magellanic Cloud,
 a common approach in both stellar and population synthesis models is to assume a linear relationship between the original $Y$ and $Z$, 
\begin{equation}
\label{eq:dydzlin}
 Y = Y_\mathrm{P} + \frac{\Delta Y}{\Delta Z} \times Z \quad
\end{equation}
where $Y_\mathrm{P}$ is the primordial helium content, i.e. the result of the Big Bang nucleosynthesis, and $\Delta Y / \Delta Z$ the ratio which provides the stellar nucleosynthesis enrichment. 

In the last four decades there has been a continuous effort to try to constrain both the $Y_\mathrm{P}$ \citep[see
e.g.][]{peebles66,churchwell74,peimbert74,peimbert76,lequeux79,kunth83,pagel86,kunth86,pagel92,mathews93,
izotov94,olive95,izotov97,olive97,peimbert02,izotov07,peimbert07,spergel07,dunkley09}
 and the ratio $\Delta Y / \Delta Z$ \citep{faulkner67,perrin77,lequeux79,peimbert80,peimbert86,pagel92,renzini94,
fernandes96,pagel98,jimenez03,izotov04,fukugita06,casagrande07}.

Indeed, as previously mentioned, the  relationship between $Y$ and $Z$ adopted in stellar models directly affects some important quantities of stellar systems, both resolved and not, inferred by comparing observations and theoretical predictions. Thus, a precise determination of $Y_\mathrm{P}$ and $\Delta Y / \Delta Z$ is of paramount importance for the studies not only of stellar evolution, but also of galaxy evolution. Furthermore, an accurate estimate of $Y_\mathrm{P}$ is of great cosmological interest, as it constrains the early evolution of the
Universe, when the Big Bang nucleosynthesis occurred. In this paper we will focus on the value of the $\Delta Y / \Delta Z$ ratio. 

In the past, several techniques have been used to determine such a ratio by means of HII regions, both galactic and extragalactic, planetary nebulae (PNe),  the Sun and chemical evolution models of the Galaxy
 (see Sect. \ref{sec:othermeth} for a brief summary of these results).

An alternative and well established way to determine the value of the helium-to-metallicity enrichment ratio takes advantage of the dependence  of the location of stars in the Hertzsprung-Russell (HR) diagram on their helium content. From the study of stellar populations in the galactic bulge \citet{renzini94} inferred $2 \leq \Delta Y / \Delta Z \leq 3$. A frequently adopted approach relies on the analysis of the fine structure of the low-MS of the local field stars in the HR diagram. Pioneers of such an approach have been Faulkner, who found  $\Delta Y / \Delta Z = 3.5$  \citep{faulkner67} and
Perrin and collaborators, who obtained $\Delta Y / \Delta Z = 5$ \citep{perrin77}. Following these early studies, \citet{fernandes96}
constrained the value of $\Delta Y / \Delta Z$ to be larger than 2, by comparing the broadening in the HR diagram of the low-mass MS stars in the solar neighborhood and the theoretical ZAMS of several $Y$ and $Z$. With a similar approach but taking advantage of Hipparcos data, 
\citet{pagel98} inferred $\Delta Y / \Delta Z = 3 \pm 2 $. This kind of approach culminated recently in the works by \citet{jimenez03} and  \citet{casagrande07} who provided, respectively, $\Delta Y / \Delta Z = 2.1 \pm 0.4 $  and 2.1 $\pm$ 0.9. 

The present analysis deals with the determination of the $\Delta Y / \Delta Z$ ratio by means of the comparison between the local K dwarf stars,
 for which accurate measures of both the [Fe/H] and parallaxes are available, and state-of-the-art stellar models. A great effort has been 
devoted to discuss the effect of the main uncertainties still present in stellar models on the inferred value of $\Delta Y / \Delta Z$. 

In Sect. \ref{sec:models} we present the set of low-mass stellar models we have computed for this paper; in Sect. \ref{sec:dataset} we describe the data set we have used; Section \ref{sec:method} contains the description of the analysis method, while Sect. \ref{sec:uncert} deals with the possible sources of uncertainty that could affect the method itself. Results for the adopted data set are presented in Sect. \ref{sec:results}. In Sect. \ref{sec:nonlin} we investigate the possibility of a non linear relation between $Y$ and $Z$. In Sect. \ref{sec:Hyatest} we apply our method to an independent and unevolved set of stars, i.e. the Hyades low-main sequence. We compare our results to those obtained by other authors, with independent methods, in Sect. \ref{sec:othermeth}. Section \ref{sec:concl} contains the final discussion and summary of the whole paper.

\section{The models}
\label{sec:models}

The stellar models have been computed on purpose for the present work with an updated version of the FRANEC evolutionary code which includes the state-of the art input physics \citep[see e.g.][]{chieffi,deg08,valle09,tognelli10}.
The main updating of the code with respect to previous versions include the 2006 release of the OPAL Equation of state (EOS)
 \footnote{http://www-phys.llnl.gov/Research/OPAL/EOS$\_2005/$} \citep[see also][]{rog96} and, for temperatures higher than
12000 K, radiative opacity tables\footnote{http://www-phys.llnl.gov/Research/OPAL/opal.html} \citep[see also][]{ig96}, while the opacities by \citet{ferg} are adopted for lower temperatures \footnote{http://webs.wichita.edu/physics/opacity}.
The electron-conduction opacities are by \citet{shtern} \citep[see also][]{pot}. The opacity tables have been calculated by assuming the solar mixture by \citet{aspl05}.

The extension of the convectively unstable regions is determined by using the classical Schwarzschild criterion. The mixing length formalism \citep{bohm58} is used to model the super-adiabatic convection typical of the outer layers. 
As it is well known, within this simplified scheme, the efficiency of convective transport depends on a free parameter that has to be calibrated using observations. We adopted the usual ''solar'' calibration of the $\alpha$ of the mixing-length. More in detail, this means that we chose the value of $\alpha = 1.97$ provided by a standard solar model (SSM) computed with the same FRANEC code and the same input physics 
we used to compute all the other stellar tracks.   

Note that the ''solar'' calibrated value of the $\alpha$ parameter strongly depends on the chosen outer boundary conditions
 needed to solve the differential equations describing the inner stellar structure, that is, the main physical quantities at the base of the
 photosphere \citep[e.g.][]{montalban04,tognelli10}.
In order to get these quantities, we followed the procedure commonly adopted in stellar computations, which consists in a direct
 integration of the equations describing a 1D atmosphere in hydrostatic equilibrium and in the diffusive approximation of radiative
transport, plus a grey T($\tau$) relationship between the temperature and the optical depth. The classical semiempirical T($\tau$) relationship 
by \citet[][]{krishna66} has been chosen. If a non-grey and more realistic model atmosphere is used, the solar calibrated value of $\alpha$ is
different \citep[see e.g.][for a detailed discussion]{tognelli10}. 

Moreover, the ''solar'' calibrated value of the $\alpha$ parameter depends also on the input physics adopted in the SSM computation. 
Thus, to the sake of consistency, if the solar calibration approach is followed to fix the $\alpha$ parameter of a set of stellar models, 
 the input physics and boundary conditions adopted to compute these models have to be the same as those used in the reference SSM. 

In spite of its widespread use, the solar calibration of the mixing-length  does not rely on a firm theoretical ground, since there are not 
compelling reasons to guarantee that the efficiency of superadiabatic convective transport should be same in stars of different mass and/or in 
different stages of evolution \citep[see e.g.][and references therein]{montalban04}. However, for what concerns the present paper, such an
approach should be quite safe, since we deal with stars in the mass range 0.7-0.9  $M_{\sun}$ and that are on the Main Sequence. 

We also computed models with a different value of $\alpha$, namely 2.4, in order to quantify the effect of an uncertainty 
in the efficiency of the mixing-length on the inferred $\Delta Y / \Delta Z$ ratio.

Our reference theoretical models have been transformed from the $(\log T_\mathrm{eff}, \log L/L_{\sun})$ to the $(B - V, M_\mathrm{V})$ diagram by means of synthetic photometry using the spectra database GAIA v2.6.1 calculated from PHOENIX model atmospheres \citep{brott}. 
We performed additional simulations adopting the \citet{castelli03} model atmospheres, to check the effect of the adopted color transformations on the inferred $\Delta Y / \Delta Z$ ratio. 

The original helium abundance in the stellar models has been obtained following Equation (\ref{eq:dydzlin}), where $Z$, once a solar mixture is assumed, is directly related to the observable [Fe/H] by
\begin{equation}
\label{eq:zeta}
 Z = \frac{1-Y_\mathrm{P}}{1+\frac{\Delta Y}{\Delta Z}+ \frac{1}{(Z/X)_{\sun}} \times 10^{-\mathrm{[Fe/H]}}} \quad 
\end{equation}
We used for $Y_\mathrm{P}$ the value 0.248, from \citet{izotov07} and \citet{peimbert07}, and for $(Z/X)_{\sun}$, the solar
metals-to-hydrogen ratio, 0.0165 from \citet{aspl05}.

Stellar models have been calculated for 9 $\Delta Y/ \Delta Z$ values (0.5, 1, 2 ... 8) and 5 [Fe/H] values (from -0.6 to +0.2 in steps of 0.2 dex). For each of these 45 combinations we calculated evolutionary tracks for 11 stellar masses (from 0.5 to 1.0 $M_{\sun}$ in steps of 0.05 $M_{\sun}$) in order to build zero age main sequence (ZAMS) curves that cover the whole HR region
corresponding to the adopted data set (see Sect. 3).
For each stellar mass we calculated its evolution starting from the pre-main sequence (PMS) phase.

To determine the ZAMS position we used the following operative criterion: we calculated the local Kelvin-Helmoltz timescale for each model, i.e.
$t_\mathrm{KH} = |\Omega| / L$, where $\Omega$ and $L$ are the gravitational binding energy and the total luminosity; then we compared this number with the local evolutionary timescale, $t_{\mathrm{ev}} = \left( \frac{1}{L}\frac{\mathrm{d}L}{\mathrm{d}t} \right)^{-1}$, i.e. the inverse of the  instantaneous rate of change of the luminosity. We found that a good operative definition of ZAMS is obtained by taking the first model for which $t_\mathrm{KH} < 100 \times t_{\mathrm{ev}}$.

As previously mentioned, an additional set of models, with the related ZAMS, has been computed for a value of the mixing-length parameter, $\alpha = 2.4$.  

Thus, the present analysis can rely on a very fine grid of stellar models consisting of about a thousand evolutionary tracks calculated from the PMS phase to the central hydrogen exhaustion.


\section{The data set}
\label{sec:dataset}
   \begin{figure}
   \centering
	\resizebox{\hsize}{!}{\includegraphics{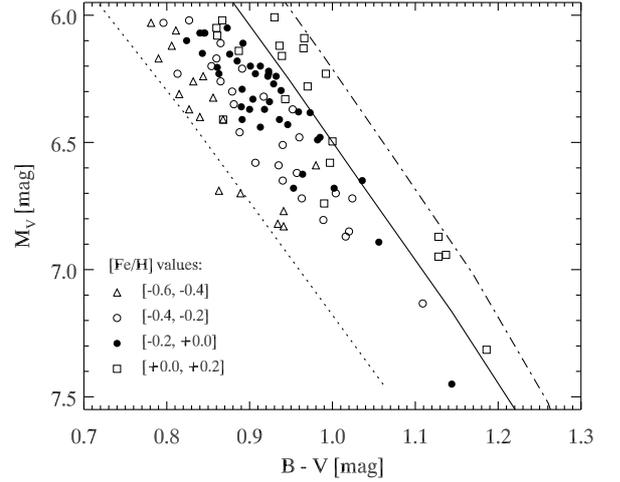}}
   \caption{The data color-magnitude diagram. Overplotted are three ZAMS all computed with $\Delta Y / \Delta Z = 2$ and [Fe/H] values of -0.6 (dotted), 0.0 (solid), 0.2 (dash-dotted).}
    \label{fig:data}
   \end{figure}

The stars of our sample have been selected among the HIPPARCOS \citep{HIP} stars with relative error on the parallax less than $5 \%$. $B$ and $V$ band photometry are also taken from HIPPARCOS data set and they have typical errors between 0.01 and 0.02 mag; combining them gives $(B -V)$ colors with errors of the order of 0.03 mag. From the parallax and the observed $V$ magnitude we computed the absolute magnitude $M_V$; with the quoted typical errors on  $V$ and parallax, the error on the absolute magnitude is $\sim 0.1$ mag, largely dominated by the error on the parallax.
Given the small values of the distances, always less than 30 pc for the stars in our sample, we also assume that the reddening is negligible, so that $(B - V)_0 =(B - V)$.
Our sample has been restricted only to those stars with $M_V~\ge~6$ in order to take objects with long evolutionary time scales and hence minimize evolutionary effects. 

[Fe/H] determinations are taken from the Geneva-Copenhagen survey of the Solar neighborhood catalog \citep[hereafter N04]{Nordstrom04}, for 86 objects, and from the catalog by Taylor \citep[hereafter T05]{Taylor05} for 21 objects. 
Of these 107 objects, 4 have [Fe/H] determinations from both catalogs; in these cases we only count the stars once, using their T05 metallicities. The total number is then reduced to 103 stars.
N04 metallicity values are derived by a calibrated relation between Str\"{o}mgren photometry measurements and spectroscopic determinations of metal abundances for a subset of objects.
The T05 catalog is instead a collection of spectroscopic determinations of metal abundances from the literature, where different results  are put by the author on the same temperature scale; if more determinations are available for the same object, they are weight-averaged according to their quality. 
As pointed out in \cite{Taylor05}, determination of [Fe/H] from different authors may suffer strong systematic deviations, due to the different temperature scale chosen. The author showed that it is possible to reach a very good zero-point accuracy  when data from different sources are put together in an appropriate way.
Indeed a systematic difference in metallicity can be seen for the 4 stars that we have in common in our subsamples of the N04 and T05 catalogs. The metallicities for these objects are shown in Table \ref{tab:met4}. There we also show the new determinations of Geneva-Copenhagen metallicities using an improved calibration \citep[hereafter H07]{Holmberg07}. As shown by the authors, the N04 and H07 spectro-photometric calibrations give internally consistent results; only one of the 4 stars in common shows a small change in the [Fe/H] between the two catalogs. The average shift in [Fe/H] on this small sample and its standard deviation is $<\mathrm{[Fe/H]}_{\mathrm{T05}} - \mathrm{[Fe/H]}_{\mathrm{N04}}> = -0.084 \pm 0.021\, \mathrm{dex}$. \cite{Taylor05} calculates the expected offset between his metallicity scale and the \cite{Nordstrom04} one; according to his Table 10, for all the stars in our N04 sub-sample this offset is expected to be $-0.023 \pm 0.017 \,\mathrm{dex}$, somewhat lower than what we find for our stars in common, which, anyway are only 4. We decided to apply the -0.023 offset to the Geneva-Copenhagen metallicities in their old version, i.e. the N04, since it is for the \cite{Nordstrom04} calibration that \cite{Taylor05} evaluates the zero-point offset. 

In the Geneva-Copenhagen survey the repeated radial velocity measurements allow detections of almost all the possible binaries in the sample; we flagged out all the suspect binaries in the catalog. The \cite{Taylor05} [Fe/H] catalog is instead free of such contaminants, given its spectroscopic nature.

Our final sample is made of 103 stars with good parallaxes, photometry and [Fe/H] determinations. The typical errors are given by: $\sigma (M_\mathrm{V}) \simeq 0.1 \,\mbox{mag}$, $\sigma (B -V) \simeq 0.03 \,\mbox{mag}$ and $\sigma(\mathrm{[Fe/H]}) \simeq 0.1 \,\mbox{dex}$. The error on the absolute magnitude is mainly due to the error on the parallax.

Metallicities range from [Fe/H] = -0.6 dex to +0.2 dex; magnitudes range from $M_V = 6.0$ mag to $ M_\mathrm{V} = 7.5$ mag. The color-magnitude diagram (CMD) for the data is shown in Fig. \ref{fig:data}, where the data are grouped in [Fe/H] bins 0.2 dex wide.

\begin{table}
\caption{[Fe/H] for the 4 stars in common in the \cite{Taylor05} and Geneva-Copenhagen catalogs. T05, N04, H07 refer to \cite{Taylor05}, \cite{Nordstrom04} and \cite{Holmberg07} respectively.}             
\label{tab:met4}      
\centering                     
\begin{tabular}{ l c c c} 
\hline
\hline                 
 HIPPARCOS ID & T05 & N04 & H07  \\
\hline                        
     HIP 79190   &  -0.410   &    -0.340  &   -0.300 \\
     HIP  88972    &  -0.157   &    -0.080  &  -0.080 \\
     HIP 99711   &  -0.006   &    0.070    &    0.070 \\
     HIP  116745  &  -0.336 &     -0.220   &    -0.220 \\
\hline
\end{tabular}
\end{table}

\section{The method}
\label{sec:method}

The idea to determine  the $\Delta Y / \Delta Z$ ratio using the position of low-mass MS stars in the HR diagram relies on the well known
 dependence of their luminosity and effective temperature on the original $Y$
 and $Z$. It is in fact firmly established that an increase of $Y$ at fixed
 $Z$ makes a star brighter and hotter. A decrease of $Z$ at fixed $Y$ leads to
 the same result. Such a behavior is the consequence of the effect on the
 opacity and mean molecular weight, i.e. the former gets higher as $Z$ increases, while the latter grows with $Y$.

As early shown by \citet{faulkner67}, who studied the effect of chemical composition variations on the position of theoretical ZAMS, a simple but instructive explanation of this behavior can be obtained by means of homology relations \citep[see also][]{fernandes96}. Within this framework, it can be shown that varying $Y$ and $Z$ in such a way that $\Delta Y \approx$ 5 $\Delta Z$, leaves the bolometric magnitude $M_\mathrm{bol}$ of ZAMS at fixed effective temperature unchanged. This explains why the broadening of the local low-MS provides a $\Delta Y/\Delta Z$ indicator.

However, this is not the whole story, as clearly proven by \citet{castellani99}, who computed a fine grid of full evolutionary models of
 ZAMS stars with several metal and helium abundances.
They showed that the  $M_\mathrm{bol}$ of ZAMS at a given $T_\mathrm{eff}$ depends quadratically on $\log Z$ and that only in a narrow range around the solar metallicity such a dependence can be reasonably linearized to $\Delta Y/\Delta Z$= 5 \citep[see also][]{fernandes96}.  
In addition, when comparing ZAMS models with real stars, one has to take into account the effect on the color indices.  \citet{castellani99} showed that a $\Delta Y /\Delta Z \approx$ 7 is required to keep unchanged the $B-V$ color at $M_V$=6 in a narrow range around $Z_{\sun}$. The discrepancy between the $\Delta Y /\Delta Z$ required to keep  the $M_\mathrm{bol}$ and $B-V$
 unchanged clearly proves the crucial role played by the model atmospheres.
 In Sect. \ref{subsec:atm} we will further discuss the effect of different assumptions on the color-transformations on the inferred $\Delta Y /\Delta Z$. 

 The use of ZAMS models is allowed as long as observational data \emph{really} lie on the ZAMS, or very close to it. Although this
 assumption is implicit in many studies which derive the $\Delta Y / \Delta Z$ from the fine structure of the low-MS, a detailed discussion on the 
effect of a deviation from such an assumption on the inferred value of the helium-to-metal enrichment ratio is still lacking.
The reason is that very faint ($M_V \ge$6) local MS stars are usually considered \emph{as if} they were still on the ZAMS, since their evolutionary timescales are longer than the Galactic Disk age. We will show in Sect. \ref{subsec:agediff} that this assumption is indeed critical, since underestimating the effects of age on both the position of the stars in the CMD and the diffusion of heavy elements below the photosphere leads to a severe bias in the final estimate of the enrichment ratio.  
We take this bias into account when giving the final result on $\Delta Y / \Delta Z$ for our work. 

   \begin{figure*}
   \centering
	    \includegraphics[width=17cm]{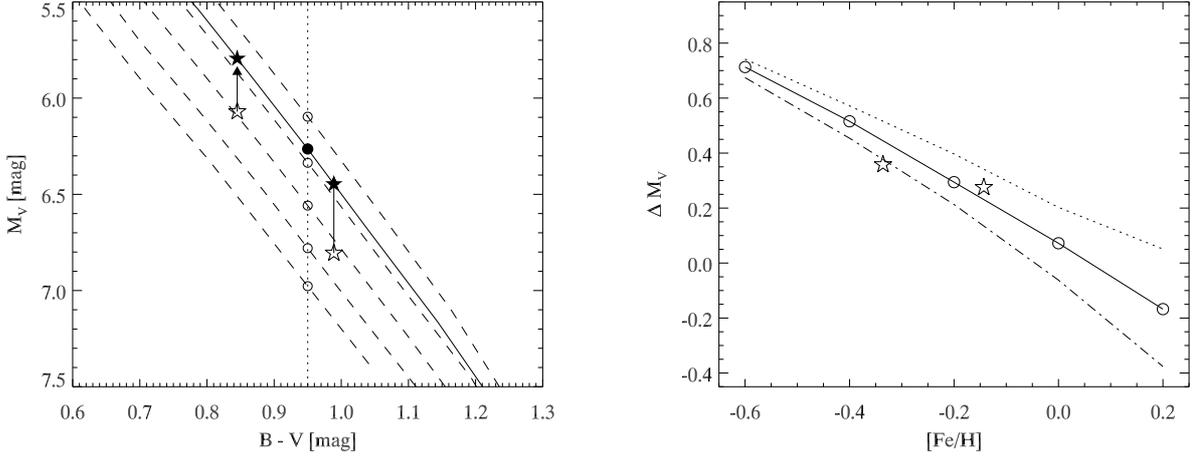}
   \caption{Illustrative example of how \emph{theoretichal} and
     \emph{observational} differences are computed and used to produce a
     $\Delta M_V$ vs. [Fe/H] diagram. \emph{Left:} Theoretical differences
     are obtained with respect to a reference ZAMS computed using $\Delta Y /
     \Delta Z = 2$ and [Fe/H] = 0.0 dex (solid line) at the reference color
     of $B - V = 0.95$ (dotted vertical line); the dashed lines are ZAMS
     computed for $\Delta Y / \Delta Z = 4$ and $\mathrm{[Fe/H]} \in
     [-0.6,0.2]$. Observational differences (represented by the distance
     between filled and empty stars) are calculated for an observed star at a given color with respect to the same reference ZAMS. \emph{Right:} The differences are used to place both ZAMS and stars in the $\Delta M_V$ vs. [Fe/H] diagram. The solid line corresponds to the curve of $\Delta Y / \Delta Z = 4$ (the ZAMS of the left panel); the dot-dashed line to $\Delta Y / \Delta Z = 0.5$; the dotted line to $\Delta Y / \Delta Z = 8$.}
    \label{fig:differenze}
   \end{figure*}

Following \citet{jimenez03}, we did not directly use the broadening of the local low-MS in the CMD, instead, we compared models and data in a diagram like that of Fig. \ref{fig:differenze}, right panel. After choosing a reference ZAMS, \emph{theoretical} differences in magnitude, $\Delta M_V$, between that reference ZAMS and the other ZAMS curves, computed for different values of [Fe/H] and $\Delta Y /\Delta Z$, are measured at a fixed value of the color index $B - V$ (see Fig. \ref{fig:differenze}, left panel).
We checked that, within the current accuracy of the data, the derived $\Delta Y/\Delta Z$ value it is not affected by changing the reference ZAMS and/or the color index value. In fact, the ZAMS loci, in the range of magnitudes and colors that is involved here, are almost parallel to each other and the effect of the uncertain position of the star caused by the observational errors is much larger then that caused by a different choice of the reference ZAMS and/or the color index value.

The differences $\Delta M_V$ obviously depend on the chemical composition, i.e. on both $\Delta Y/\Delta Z$ and [Fe/H],
 as is clearly visible in Fig. \ref{fig:differenze} (right panel). 
\emph{Observational} differences between the data set and the same reference ZAMS are also measured in the way illustrated in Fig. \ref{fig:differenze} (left panel) and are plotted in Fig. \ref{fig:differenze} (right panel) as a function of [Fe/H].

To find the value of $\Delta Y / \Delta Z$ that gives the best fit to the data, we assign to each star errors in the three quantities $M_V$, $B - V$ and [Fe/H]. The errors in $M_V$ and $B - V$ are assigned in the CMD, i.e. \emph{before} the differences $\Delta M_V$ are calculated; then the error in [Fe/H] is assigned in the $(\Delta M_\mathrm{V}, \mathrm{[Fe/H]})$ diagram.
Magnitude and color errors are considered to be distributed as gaussian with $\sigma$ equal to the quoted uncertainty for that star; the assumption of gaussian errors is reasonable, considering that they come from the HIPPARCOS photometric errors plus (for the absolute magnitude) the HIPPARCOS parallax error; these two sources of error are independent and our objects are all close by and have well determined parallaxes, so that they don't suffer the Lutz-Kelker bias. Regarding the [Fe/H] value the situation is different, since the errors associated to each value are highly affected by systematic effects like the choice of the temperature scale; for this reason, and since we can not reconstruct the real error distribution in [Fe/H], we adopted an uniform distribution of [Fe/H] errors. Anyway, we have also checked that a gaussian distribution for [Fe/H] values leaves the results essentially unaffected.

\begin{figure}
   \centering
\resizebox{\hsize}{!}{\includegraphics{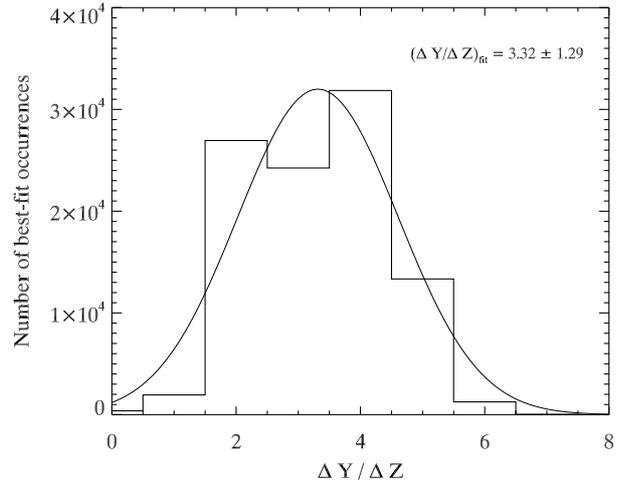}}
   \caption{Results of $10^5$ simulation runs to determine the best fitting $\Delta Y / \Delta Z$ value for our 103 stars data set, according to Equation (\ref{eq:somma}). The best-fitting gaussian is also shown, together with its mean value and standard deviation.}
    \label{fig:results_all}
 \end{figure}

Once a ''new'' data set is created from the original value plus the errors, we determine the theoretical curve which minimizes the quantity:
\begin{equation}
\label{eq:somma}
	\mathcal{D}_j = \sum_i \left[\frac{\Delta M_{V,i} - \Delta M_{V,j}(\mathrm{[Fe/H]}_i)}{\sigma (\Delta M_{V,i})}\right]^2   \; ,
\end{equation}
where $j$ runs over the curves (i.e. over different $\Delta Y/\Delta Z$ values) and $i$ over the data. The quantity $\Delta M_{V,j}(\mathrm{[Fe/H]}_i)$ is the value of the magnitude difference for the $j$-th theoretichal ZAMS, calculated for the [Fe/H] value of the $i$-th star.
The scheme is iterated in order to obtain not only the best fitting value but also an estimate of the error on the final result due to the observational uncertainties. 
For each iteration we take the value that minimizes Equation (\ref{eq:somma}) and then plot an histogram of the number of  best-fit occurrences for each $\Delta Y / \Delta Z$ in our models grid.
A gaussian fit to the histogram is performed and the mean and the standard deviation of this gaussian are used as estimators of the true $\Delta Y / \Delta Z$ and its error.
Figure \ref{fig:results_all} shows the histogram of occurrences for a Monte Carlo simulation with a total of $10^5$ iterations using our data set of 103 stars; superimposed is the best-fitting gaussian. We will discuss in Sect. \ref{sec:results} why, in this particular case, a gaussian curve does not fit well the histogram of occurrences, with the central part of the histogram which is too broad and flat to be well approximated by a single gaussian. We just anticipate that this discrepancy is mainly due to the offsets in the [Fe/H] scales for N04 and T05, even after the correction of Table 10 of \cite{Taylor05} is applied to the N04 data.

\section{Analysis of possible sources of uncertainty using artificial data sets}
\label{sec:uncert}

To check the reliability of the procedure described above and to evaluate the contribution of the various possible sources of uncertainty,
 we applied our method to a number of artificial data sets with controlled input parameters.

These sets have been created by interpolation in our fine grid of stellar models.
Stellar masses are randomly generated from a power law initial mass function (IMF) $\frac{\mathrm{d}N}{\mathrm{d}m} = c \times m^{-\alpha}$ with a slope of $\alpha = 2.3$ \citep{Salpeter55,Kroupa01}; an IMF with a unique value of the exponent is a good law in our range of simulated masses ($0.5~\le~M_\mathrm{sim}/M_{\sun}~\le~1$).
The chemical composition is calculated by fixing the input value of $\left(\Delta Y/\Delta Z\right)_\mathrm{in}$ and extracting random values for [Fe/H] for each star; $Y$ and $Z$ are then calculated using the two Equations (\ref{eq:dydzlin}) and (\ref{eq:zeta}). 
In what concerns the stellar ages generation, we performed different kind of simulations, adopting three age-laws (i.e. star formation rates, SFRs), namely a Dirac's delta centered about a given age (i.e. coeval stars), a uniform and an exponentially decaying age distribution, respectively, in the range 0-7 Gyr, a reasonable approximation for the age of the galactic disk.

Since the FRANEC code includes a treatment of diffusion of the elements, given a star of age $\tau$, we take from the models the corresponding surface value of [Fe/H]$_\tau$, which is different from the initial value [Fe/H]$_0$ because of diffusion itself. This is the value that we use for the recovery, because in a real star the observed [Fe/H] is the \emph{present} one and not the initial. Note that neglecting diffusion may lead to a bias (an underestimate, indeed) in the final $\Delta Y / \Delta Z$ if the sample of stars is old enough to have experienced a not negligible amount of diffusion of heavy elements; we will show this in Sect. \ref{subsec:agediff}.

When generating the stellar models parameters for our sets of stars, we do not take into account any age-metallicity relation, hence [Fe/H] and age values are independently and randomly extracted. Recent works \citep{Nordstrom04,Holmberg07} have indeed shown that there is no evidence of an age-metallicity relation for local disk stars. 
Given mass, age and chemical composition, we interpolated in our fine grid of stellar models to obtain the observational properties of the 
simulated stars.
The number of stars in each simulated set is 110, a number comparable to the 103 stars of our real data sample.

In the following and in the related figures, we will refer to the input parameters of our simulated samples using the subscript \emph{in} and to the output of the recovery method using the subscript \emph{out}.

\subsection{The effect of measurements errors}
\label{subsec:measerr}
The first test we have performed was made to check whether our recovery method was able to get the right $\Delta Y / \Delta Z$ from  an ''ideal'' sample of stars affected only by observational errors on the magnitude, color and [Fe/H] values. By ideal we mean a sample that, regardless of masses and [Fe/H] distribution (which indeed were generated in a completely random fashion), contains only stars \emph{really lying on the ZAMS}. 
It is worth to point out that in the case of the real data, this is only a simplifying assumption, which can not be exactly fulfilled, since the
 observed stars in our sample have unknown ages that span the whole range of ages in the Galactic disk.

Once the artificial sample has been generated, we associated to each star an error in absolute magnitude, color and [Fe/H] typical of our real sample of data, i.e., 0.1 mag, 0.03 mag and 0.1 dex respectively (see Sect. \ref{sec:dataset}) .

\begin{figure}
   \centering
\resizebox{\hsize}{!}{\includegraphics{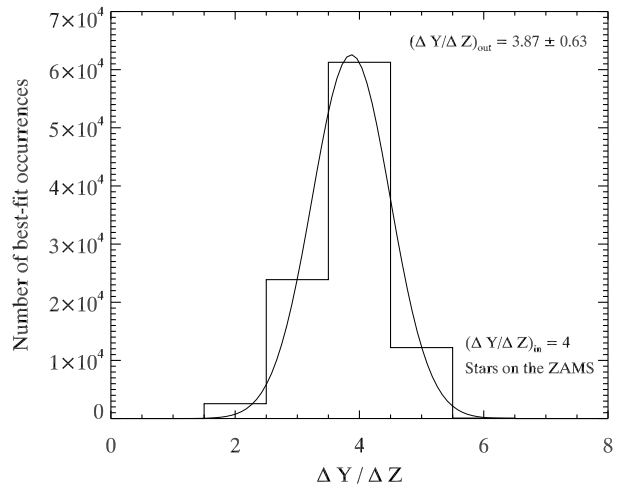}} 
   \caption{Results of a test for a sample of 110 simulated stars lying on the ZAMS (see text for more details). Overplotted is the best-fit gaussian.}
    \label{fig:results_finto_noage_unito}
 \end{figure}

We found that our recovery method is not affected by observational errors of this order of magnitude.
 As it is possible to see in Fig. \ref{fig:results_finto_noage_unito}, given a
 $\left( \Delta Y/\Delta Z \right)_\mathrm{in}$ of 4, the best value that
 comes out from our Monte Carlo method and the gaussian fit to the histogram of occurrences is indeed $\left( \Delta Y/\Delta Z
 \right)_\mathrm{out} = 3.87 \pm 0.63$. So the outcome of the method is perfectly consistent with the input value and; moreover it has a very small range of 1.26 at a level of $1\sigma$, which is comparable to the
 resolution in our models grid (1 unit), and which we may quote as the nominal or intrinsic error of the method, associated to the typical error of the actual data.

\subsection{The effects of age and heavy elements diffusion}
\label{subsec:agediff}
\begin{figure}
   \centering
\resizebox{\hsize}{!}{\includegraphics{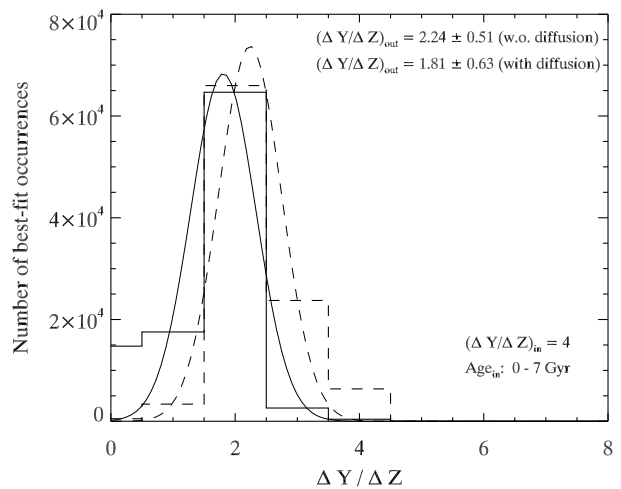}}
   \caption{Results of a test for a sample of 110 simulated stars not lying on the ZAMS. Ages are uniformly distributed between 0 and 7 Gyr. The solid line gives the result when heavy elements diffusion is taken into account, while the dashed line corresponds to the case where [Fe/H]$_t$ is equal to [Fe/H]$_{ZAMS}$. Overplotted are the best-fit gaussians.}
    \label{fig:agebias1}
 \end{figure}

Although our stellar sample has been obtained by selecting very faint stars ($M_V > 6$ mag, i.e. $M \lesssim 0.9 M_{\sun}$, the actual value depending on the chemical composition), we found that evolutionary effects strongly affect the final result introducing a non negligible \emph{bias}. This is actually one of the most important results of this work. Thus, one should be very careful in properly taking into account the evolutionary effects, i.e. the displacement from the ZAMS, when helium-to-metals enrichment ratio is derived from the low MS fine structure. 

Indeed, by creating artificial stellar data sets with $(\Delta Y/\Delta
Z)_\mathrm{in} = 4$, but no longer on their ZAMS position, we found an output
value of our recovery method of $(\Delta Y/\Delta Z)_\mathrm{out}  \sim 2$. The actual best-fit value for each simulated data set depends on the exact
parameters used to generate the artificial sample, like the maximum age or the
functional form of the age distribution (uniform or with an exponentially
decaying SFR). The total number of stars, their [Fe/H] and magnitude ranges
are always kept the same between the different simulated data set. Figure
\ref{fig:agebias1} shows the results of our method using data sets where the
simulated stars have ages uniformly distributed between 0 and 7 Gyr. The
dashed line indicates the results when the [Fe/H] values associated to each
star at a given age are the same as the ones at the ZAMS; whereas the solid
line indicates the results when diffusion is taken into account, i.e. the
value at a given time [Fe/H]$_t$ is different from [Fe/H]$_{ZAMS}$. 
Even without taking into account diffusion, the method gives an output value of $(\Delta Y/\Delta Z)_{out} = 2.24 \pm 0.51$, quite different from  $(\Delta Y/\Delta Z)_{in} = 4$. When also diffusion is taken into account, the best fit value suffers an additional shift, with $(\Delta Y/\Delta Z)_{out} = 1.81 \pm 0.63$.
Since we don't know what is the real age distribution of the stars of our sample we can not really quantify the bias, but after many experiments with several data set, we conclude that it must be of the order of $ \mathcal{B}_{age}(\Delta Y/\Delta Z) = -2 \pm 0.5$.

\begin{figure}
   \centering
\resizebox{\hsize}{!}{\includegraphics{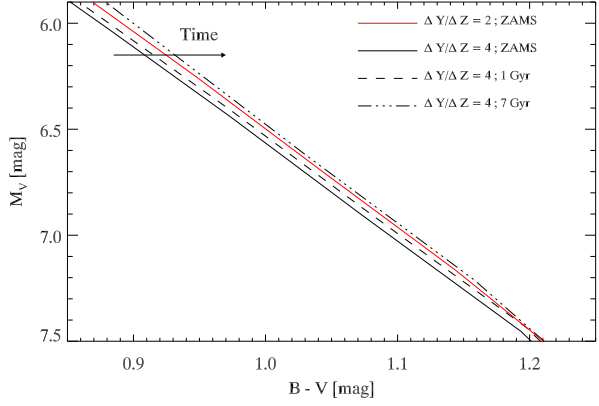}}
   \caption{The evolution of stars mimics lower values of the enrichment ratio. A 7 Gyr isochrone calculated for $\Delta Y/\Delta Z = 4$ and [Fe/H]=0 is very similar to a ZAMS with $\Delta Y/\Delta Z = 2$ and [Fe/H]=0. The arrow indicates the effect of evolution on the position of the isochrones.}
    \label{fig:Iso+ZAMS}
 \end{figure}

The effect of evolution on the derived enrichment ratio is easy to understand by looking at Fig. \ref{fig:Iso+ZAMS}. Here two ZAMS with different values of $\Delta Y/\Delta Z$, namely 2 and 4, and same [Fe/H]~=~0.0 are shown together with isochrones of 1 and 7 Gyr, calculated with  $\Delta Y/\Delta Z = 4$ and [Fe/H] = 0.0; it is clear that evolution causes a shift of the whole curve towards redder colors\footnote{A single star becomes hotter and more luminous after leaving the ZAMS, here we are referring to the \emph{overall shape} of the isochrones and ZAMS in this range of magnitudes and colors.} in a completely indistinguishable fashion as a lower $\Delta Y/\Delta Z$ does.

As a summary, this means that, even if our \emph{observational} data set has been selected with a very strict cutoff of $M_V = 6$ mag, evolutionary effects still play an important role. The real, unbiased value of $\Delta Y/\Delta Z$ coming out from our analisys has then to be
 corrected, by subtracting $\mathcal{B}_{age}$ from the nominal value given by the Monte Carlo method. We then expect that the real value is higher by about two units than what can be found by blindly applying this method to the data.

\subsection{The effects of the uncertainty on the mixing-length efficiency}

\begin{figure}
   \centering
\resizebox{\hsize}{!}{\includegraphics{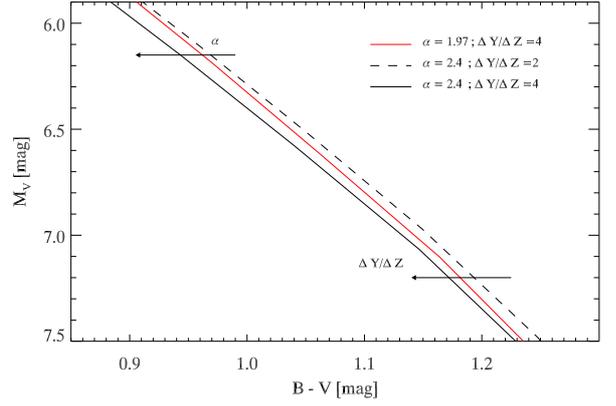}}
   \caption{Effect of changing the mixing length parameter from $\alpha = 1.97$ to 2.4 on the calculated ZAMS; all the models shown are calculated for [Fe/H] = +0.2 dex. The arrows indicate in which direction the ZAMS move in the CMD when increasing $\alpha$ (upper arrow) or $\Delta Y/\Delta Z$ (lower arrow).}
    \label{fig:twoalpha}
 \end{figure}

The current generation of stellar models is not yet able to firmly predict the effective temperature of stars with a convective envelope, such as those belonging to our sample. The reason is that a satisfactory and fully consistent theory of convection in superadiabatic regimes is still lacking, hence a very simplified approach is usually followed. The approach commonly adopted in the vast majority of evolutionary codes is to
implement the mixing-length theory \citep{bohm58}, in which the average efficiency of convective energy transport depends on a free parameter $\alpha$ that must be calibrated. 
Our reference set of stellar models has been computed adopting our solar calibrated value of the mixing length parameter, namely $\alpha = 1.97$.   Nevertheless, we calculated a whole new grid of models using $\alpha = 2.4$ in order to evaluate the effect of this still uncertain parameter on the derived value of $\Delta Y/\Delta Z$. As well known, the predicted effective temperature of a stellar model with a convective envelope is an increasing function of the value of the mixing length parameter $\alpha$, as a consequence of the shallower temperature gradient due to a more efficient convective energy transfer. 

Figure \ref{fig:twoalpha} shows the effect of different adopted $\alpha$ values on the calculated ZAMS. The assumed mixing-length parameter affects also the inferred $\Delta Y/\Delta Z$ ratio, since it directly influences the predicted position in the HR diagram of the ZAMS models. As one can easily see in Fig. \ref{fig:twoalpha}, in order to recover the ZAMS locus of models computed with an  higher value of $\alpha$, a lower $\Delta Y/\Delta Z$ ratio is needed. Notice also that, the impact of the mixing-length efficiency on the predicted effective temperature of ZAMS models of the same chemical composition becomes progressively smaller at faint magnitudes, i.e. for very low-masses. Such a behavior is the consequence of the almost adiabatic nature of convection in the envelopes of very-low mass stars ($M \lesssim 0.7 M_{\sun}$), characterized by high densities and low temperatures. 

\begin{figure}
   \centering
\resizebox{\hsize}{!}{\includegraphics{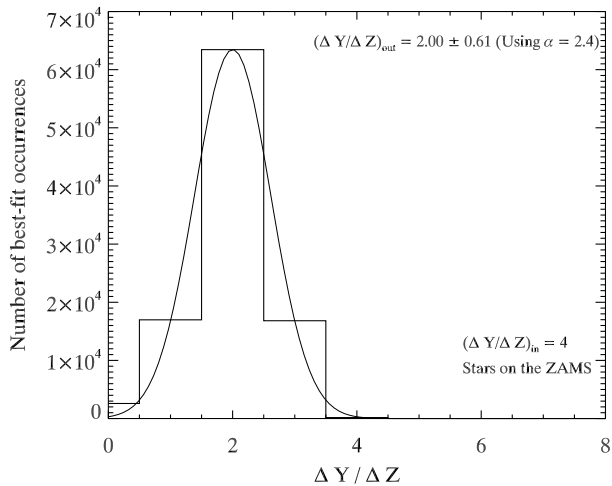}} 
   \caption{Results for the same data set of Fig.\ref{fig:results_finto_noage_unito}. In this case we ran our Monte Carlo method using ZAMS calculated with a mixing length parameter $\alpha = 2.4$. Overplotted is the best-fit gaussian.}
    \label{fig:results_finto_noage_al2p40}
 \end{figure}

Since an increase in $\alpha$ affects the models in the same direction as an increase in $\Delta Y/\Delta Z$, we find, as expected, that our recovery
method gives a lower value of the enrichment ratio when ZAMS calculated for
$\alpha = 2.4$ are used, that is, $(\Delta Y/\Delta Z)_{out} =  2.00 \pm
0.61$, quite different from  $(\Delta Y/\Delta Z)_{in} = 4$. This is shown in Fig. \ref{fig:results_finto_noage_al2p40} which reports the output of the Monte Carlo method, when the same data set of Sect. \ref{subsec:measerr} is used, i.e. a set of stars lying on the ZAMS calculated with $\alpha = 1.97$.
 We already mentioned that the use of the solar calibration should be the safer choice when dealing with Main Sequence stars of solar-like masses such those of our data set. On the other hand, this numeric experiment allows to quantify the effect of a wrong assumption of the mixing length parameter in the adopted stellar models on the inferred helium-to-metals enrichment ratio.

\subsection{The effects of different transformations from the theoretical to the observational plane}
\label{subsec:atm}
\begin{figure}
   \centering
\resizebox{\hsize}{!}{\includegraphics{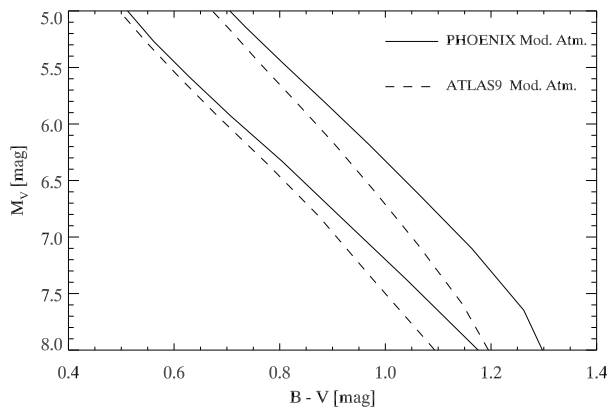}}
   \caption{ZAMS transformed into the observational plane using PHOENIX (solid lines) and ATLAS9 (dashed lines) model atmospheres. All the ZAMS shown have been calculated with $\Delta Y/\Delta Z = 4$ and [Fe/H] = -0.6 (left) or +0.2 (right).}
    \label{fig:twoatm}
 \end{figure}

To compare observational data with stellar models, one needs to transform the theoretical predictions of evolutionary codes from the $(\log T_\mathrm{eff}, \log L/L_{\sun})$ plane (HR diagram) to the observational plane, in our case the $(B-V, M_\mathrm{V})$ Color-Magnitude diagram.
A common procedure is to use synthetic stellar spectra, calculated using model atmosphere codes, and to convolve them with the filter throughput of the photometric system needed; this technique, referred to as \emph{synthetic photometry} is described in detail in, e.g. \cite{girardi02}.
We transformed our models using PHOENIX \citep{brott} and ATLAS9 \citep{castelli03} model atmospheres; the UBVRIJHKL Johnson-Cousin-Glass photometric system zero points are those of Table A1 in \cite{bessel99}.

Both model atmosphere grids completely cover our range of $T_\mathrm{eff}, \log g$ and [Fe/H]. An exhaustive description of these models is far beyond the scope of this paper, nevertheless we want to mention briefly some of the most important differences between them.
The ATLAS9 code solves the radiative transfer equation in a plane parallel atmosphere, while PHOENIX code takes into account the curvature of the atmosphere, even though in spherical symmetry, i.e. keeping a 1D approach. This difference is not very important for dwarf stars,
 for which the curvature (i.e. the ratio of the extent of the atmosphere to the radius of the base of the atmosphere) is very small. More important is the different database of molecular opacities for the two models. While atomic opacities databases are similar between the two models, PHOENIX models include a lot of molecular species ($\sim 650$) and molecular transitions ($\sim 550$ millions) which play an important role for low mass stars, specially as the metallicity increases. 

Figure \ref{fig:twoatm} shows the comparison between ZAMS models transformed using the two sets of model atmosphere. It is evident how the PHOENIX ZAMS are always redder in the $B-V$ color than the ATLAS9 ones at fixed magnitude. Moreover the differences in stellar colors increase with decreasing mass (increasing magnitude) and also, at a given magnitude, the differences increase with
 increasing metallicity (from left to right in Fig. \ref{fig:twoatm}). Note that both a lower temperature and a higher metal content favor the formation of chemical composites; in particular the first molecules start forming when the effective temperatures drops below
 $\sim 4000 K$.

The sizeable difference in magnitude and color index between the same theoretical ZAMS transformed into the observational plane by the two quoted model atmospheres, directly translates into a large difference in the inferred $\Delta Y/\Delta Z$ value. 
 We used the ATLAS9 ZAMS, running our recovery method  on a simulated data set with $(\Delta Y/\Delta Z)_\mathrm{in} = 4$ but generated using the PHOENIX ZAMS. The ATLAS9 ZAMS are so much bluer than the PHOENIX ones that in each iteration of the method we always find the lowest possible value of  $(\Delta Y/\Delta Z)_\mathrm{out}$ available in our grid of models, i.e. 0.5.

The uncertainty due to the chosen model atmosphere is then by far the most severe source of uncertainty affecting the final value of the enrichment ratio, at least among the uncertainties coming from the models side.

\subsection{The effects of different choices for the heavy elements mixture}
\label{subsec:mixt}

\begin{figure}
   \centering
\resizebox{\hsize}{!}{\includegraphics{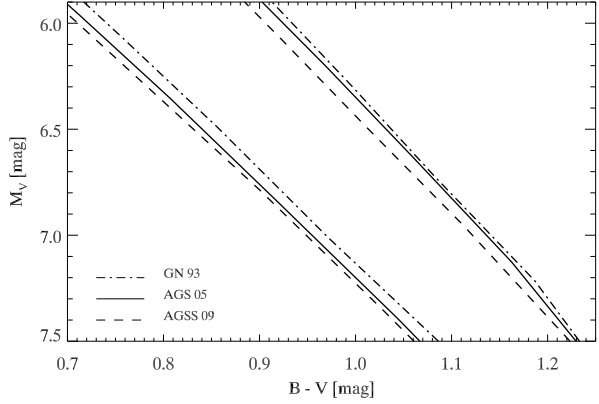}} 
   \caption{ZAMS calculated for the three different mixtures, GN93, AGS05 and AGSS09, with $\Delta Y/\Delta Z$ = 4 and [Fe/H] = -0.6 (lower B-V values) or [Fe/H] = +0.2 (higher B-V values).}
    \label{fig:Mixture}
 \end{figure}

As previously explained, all the models in our grid have been calculated using the solar-scaled mixture by \cite{aspl05}, hereafter AGS05.
A new version of the solar mixture has been recently published by the same group in \cite{aspl09}, hereafter AGSS09.
The true dependence of the inferred $\Delta Y / \Delta Z$ on the mixture choice could be
evaluated only by re-calculating an equivalent grid of models using the new
mixture and by re-running the whole procedure illustrated in this work. However,
the exact determination of the solar mixture is still an open problem, thus a
recalculation of all the models is not needed, in our opinion, until this issue
will be definitively settled. Nevertheless, to have an idea of the influence of a variation
of the solar mixture on our results, we computed two new sets of ZAMS with different mixtures and compared them with our reference AGS05 ZAMS in the CMD.
This gives at least an indication of how the inferred enrichment ratio may depend on the mixture.
In addition to the ZAMS for AGS05 and AGSS09 mixtures, we also computed and compared ZAMS
calculated with the older mixture by \cite{grev93}, hereafter GN93, still widely used in the literature.
We highlight the fact that we can control the effect of the mixture changes
on the stellar structure, by using opacity tables calculated with different
mixtures both for the high-temperature (OPAL) and low-temperature
(\cite{ferg}) opacities. On the other hand we can not evaluate the effects of
different mixtures on the model atmosphere, i.e. on the transformation of our
theoretical models from the theoretical HR diagram to the CMD, since PHOENIX synthetic
spectra are available only for a single mixture of heavy elements.

The effect of the adopted solar mixture on the models is twofold. First, for a given 
global metallicity $Z$, changing the internal distribution of metals will
mainly affect the opacity and the nuclear burning efficiency (via the CNO
abundances). For models with the same $Y$ and $Z$, the ZAMS computed with the
GN93 mixture are bluer than that with the AGSS09 one, which, in turn, is bluer than the
ZAMS computed with the AGS05 mixture \citep[see e.g.][]{deg06, tognelli10}.  
Second, for a given choice of the two values of $\Delta Y/\Delta Z$ and
[Fe/H], it is clear from the coupled equations (\ref{eq:dydzlin}) and
(\ref{eq:zeta}) that the values of $Y$ and $Z$ will be different,
 given the different values of $(Z/X)_{\sun}$. More in detail, the 
GN93 mixture provides models with higher $Z$. These models are thus redder on the CMD,
 than those calculated with AGSS09 mixture. The latter mixture, in turn, provides models that have higher $Z$ and  appear redder
 than the AGS05 mixture. 
Thus, the effects on the opacity and on the scaling, i.e. the actual value
 of $Z$ given [Fe/H] and the different $(Z/X)_{\sun}$, affect the ZAMS position
 in opposite directions.

We calculated models for the two additional mixtures by fixing $\Delta
Y/\Delta Z = 4$ and choosing the two values $\mbox{[Fe/H]} =$ -0.6 and 0.2.
The results of the calculations are shown in Fig. \ref{fig:Mixture}.
The corresponding $Y$ and $Z$ values for each mixture are shown in Table
\ref{tab:YZmix}. From Fig. \ref{fig:Mixture} one can see that 
GN93 models seem to be slightly redder than our reference AGS05 models; when
compared to observational data, an higher value of $\Delta Y/ \Delta Z$
would then be needed to reproduce observations, as compared to the AS05 mixture.
 AGSS09 ZAMS go in the opposite direction meaning that the inferred value of
 the enrichment ratio would be lower when compared to AGS05.
In the comparison between AGS05 and AGSS09, which have quite similar
$(Z/X)_{\sun}$, the effect on the opacity and burning efficiency seems to
prevail on the scaling; viceversa in the comparison between AGS05
and GN93, the huge difference in $(Z/X)_{\sun}$ compensates for the opacity
effect, and the much higher Z value in the GN93 case brings the ZAMS models
toward redder colors when compared to AGS05.

The relative behavior of different ZAMS is moreover a non trivial function of the actual value of [Fe/H],
 given the non linear relation between the two couples of values 
$(\Delta Y / \Delta Z, \mbox{[Fe/H]}) \mapsto (Y,Z)$, represented
 by the two equations (\ref{eq:dydzlin}) and (\ref{eq:zeta}).
Overall it seems that changing the mixture affects our method in a non
negligible way, but a totally consistent check could be done only when
 the appropriate model atmospheres are used to do the transformation from the
 HR diagram to the CMD and only calculating complete grids of models for
 different mixtures.

\begin{table}
\begin{minipage}[t]{\columnwidth}
\caption{$Y$ and $Z$ values for the three different adopted mixtures (see text
  in Sect. \ref{subsec:mixt}),
for two reference values of [Fe/H] and $\Delta Y/ \Delta Z = 4$. The solar
$(Z/X)_{\sun}$ values corresponding to the three mixtures are also indicated.}
\label{tab:YZmix}
\centering
\renewcommand{\footnoterule}{}  
\begin{tabular}{|c|ccc|}
\hline
\hline
    & GN93   &  AGS05  & AGSS09 \\
    &  $\left(\frac{Z}{X}\right)_{\sun} =  0.0245$  & $\left(\frac{Z}{X}\right)_{\sun} =  0.0165$ & $\left(\frac{Z}{X}\right)_{\sun} =  0.0181$  \\[6pt]
\hline
\multirow{2}{*}{[Fe/H] = -0.6} & $Y = 0.2660$     & $Y = 0.2602$    & $Y =0.2614$ \\
                                                & $Z = 0.00449 $ & $Z = 0.00305 $ & $Z = 0.00334$ \\ 
\hline
\multirow{2}{*}{[Fe/H] = 0.2} & $Y = 0.3458$  & $Y = 0.3178$  & $Y =0.3235$ \\
                                                & $Z = 0.0245 $ & $Z = 0.0174 $ & $Z = 0.0189$ \\ 

\hline

\end{tabular}
\end{minipage}
\end{table}

\section{Results using the observational data}
\label{sec:results}
After having carefully checked the capability of our recovery method and having studied  many uncertainty sources by means of controlled artificial data sets, we applied the above technique on the sample of real observational data described in Sect. \ref{sec:dataset}.

Figure \ref{fig:results_all} shows the result of the Monte Carlo recovery method applied to the local low-MS field stars, which provides a nominal value of the enrichment ratio of $\Delta Y/\Delta Z = 3.32 \pm 1.29$. 

This result for the enrichment ratio is obtained using our favorite set of
models, i.e. ZAMS calculated with the solar
 $\alpha=~1.97$ and transformed using the PHOENIX model atmospheres.
We showed in Sects. 5.3 and 5.4 the significative effects on the inferred
helium-to-metals enrichment ratio of a different choice of the mixing
length parameter $\alpha$ and of the model atmosphere, respectively, in the
case of synthetic data. The recovery procedure applied to the real
observational data provides a nominal value of the enrichment ratio of 
$\Delta Y/\Delta Z= 1.59 \pm 1.01$ adopting the set of theoretical models computed
with $\alpha=~2.4$ and transformed into the observational plane by means of
PHOENIX atmosphere models and $\Delta Y/\Delta Z= 0.5$ adopting our standard set
of models with $\alpha=~1.97$ but the ATLAS9 atmosphere models.
  
Moreover, we will show a self-consistency check in Sect. \ref{sec:Hyatest}, to prove that \emph{inside} our choice the result we obtain for $\Delta Y/\Delta Z$ can be used to calculate models that fit very well an independent data set, in particular the Hyades main sequence.

\begin{figure}
   \centering
\resizebox{\hsize}{!}{\includegraphics{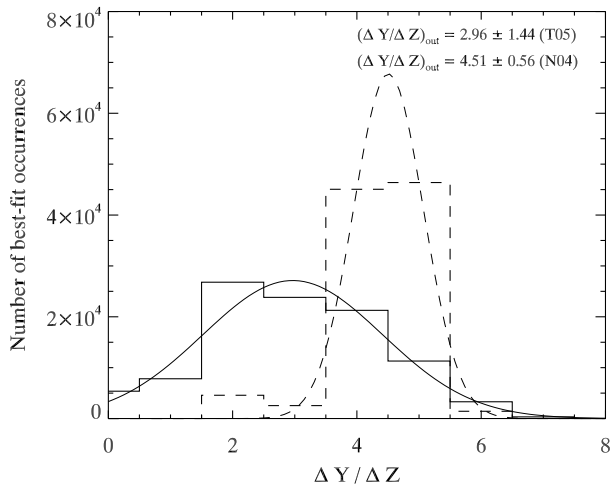}} 
   \caption{Results of our Monte Carlo method when applied separately to the subsets of objects with [Fe/H] from T05 and N04 (the latter corrected using Table 10 of T05). The solid lines are for T05, the dashed for N04. Overplotted are the best-fit gaussians.}
    \label{fig:NH_HT}
 \end{figure}

In Sect. \ref{sec:dataset} we mentioned the fact that the fit of a gaussian distribution to the histogram of occurrences of Fig. \ref{fig:results_all} is not perfect. The actual distribution looks quite flat in the range $\Delta Y/\Delta Z \in [2, 4]$ and the $\sigma$ of the fitting gaussian is twice as large than what we would expect from the observational uncertainties alone (see Sect. \ref{subsec:measerr}).
We also mentioned that our [Fe/H] estimates come from two different catalogs, and that the N04 data have been re-zeroed according to T05 transformations. Nevertheless, when considering the two subsets of data separately, a large difference in the final result is still visible, as if the different [Fe/H] scale adopted, even after the re-zeroing, would mimic a bimodal chemical composition distribution.
From Fig. \ref{fig:NH_HT} it is clear that the two results for the separate
subsets are inconsistent at $1\sigma$ level. The final result we get for the
whole set, is then an average of the two distributions of
Fig. \ref{fig:NH_HT}; unfortunately this method is quite sensitive to the adopted scale of [Fe/H] determinations. We won't choose any of the two subsets as the \emph{best one}, we just state that systematics is probably still affecting the [Fe/H] of our objects, even after the correction of N04 zero point.
Anyway, in order to proceed further with the analysis we will use the results of the gaussian fit to the histogram of occurrences for the whole sample of data as our nominal value for $\Delta Y/\Delta Z$.
As we showed in Sect. \ref{subsec:agediff}, to obtain the real $\Delta Y/\Delta Z$ value, the nominal value of $\Delta Y/\Delta Z = 3.32 \pm 1.29$ provided by the recovery method must be corrected for the age-bias, i.e. the artificial shift in the recovered value of the enrichment ratio caused by neglecting evolutionary effects and diffusion. We estimated the extent of such a correction to be of $\mathcal{B}_{age} = -2 \pm 0.5$ by means of Monte Carlo simulations. 
Thus, the final corrected value of the helium-to-metals enrichment, combining the errors in quadrature is $\Delta Y/\Delta Z = 5.3 \pm 1.4$.

\section{A possible non-linear relation between $Y$ and $Z$?}
\label{sec:nonlin}

In their recent work, \cite{casagrande07} suggested that the assumption of a linear relation between metals and helium enrichment of the ISM, which leads to the definition of the enrichment ratio itself (see Equation \ref{eq:dydzlin}) may not necessarily be valid.
We tried to check whether it is possible to unambiguously infer such a non linear behavior from  the presently available data accurateness. We performed the test by means of an artificial and controlled data set built interpolating our stellar models. The synthetic stellar sample mimics the main characteristics of the real data set. In particular, we simulated stars divided in metallicity bins 0.1 dex wide in the range [Fe/H] $\in [-0.6,+0.2]$, with the same number of stars in each bin as our observational data, for a total number of 103 objects. To avoid the additional difficulties introduced by the age-bias, the simulated stars were taken on their ZAMS position.
We then ran our Monte Carlo method bin by bin to see whether our recovery is able to recover the input value for $\Delta Y/\Delta Z$ in each bin, given the typical observational errors on our objects.

\begin{figure*}
   \centering
	    \includegraphics[width=17cm]{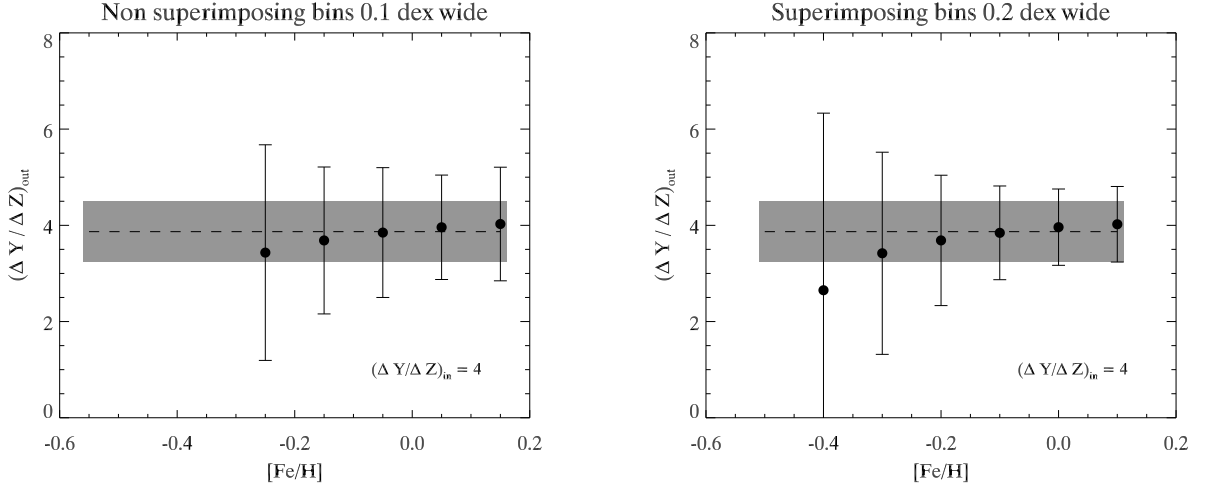}
   \caption{Output values of the recovery method when data are binned in metallicity (see text for details). The green dashed lines represent the $(\Delta Y/\Delta Z)_\mathrm{out}$ when the whole simulated sample is used. The input value is $(\Delta Y/\Delta Z)_\mathrm{in} = 4$.}
    \label{fig:binned}
 \end{figure*}

Figure \ref{fig:binned} shows the outcome of such an experiment. On the two panels each point represents the result of the Monte Carlo method when applied to the corresponding [Fe/H] bin, with its $1\sigma$ uncertainty; both values come from the gaussian fitting of the histogram of occurrences. The dashed lines represent the results of the gaussian fit the histogram of occurrences when the method is applied to the whole sample of simulated stars; the shaded area corresponds to the $1\sigma$ uncertainty around the mean. The two panels represent results for two different ways of binning the data; on the left panel we show the results when data are binned in 0.1 dex wide non overlapping bins; on the right panel the results are for  0.2 dex wide bins, in this case adjacent bins have an overlap of 0.1 dex, meaning e.g. that a [Fe/H] = -0.35 star can be found in both the [-0.5, -0.3] and the [-0.4, -0.2] bins.
Regardless from the way the binning is performed, it follows from this numerical experiment that the method works fine only for [Fe/H] values equal or greater than -0.3 dex. $\Delta Y/\Delta Z$ determinations with this method, considering only lower values of the metallicity are unreliable. The missing points in the two panel, correspond to bins where the gaussian fit failed, simply because the histogram of occurrences is too irregular to be fitted by a gaussian.

Hence any discussion on a possible non linearity of the helium-to-metals enrichment relation, extending over a wide range of metallicities, can not be carried on when $Y$ and $Z$ value are determined from low-main sequence fitting. The fine structure of the low main sequence at values of the metallicity lower than -0.3 dex is indeed too weakly dependent on the chemical composition, to be used to infer any non linear behavior of the ISM enrichment in helium and metals.
This is quite clear from the right panel of Fig. \ref{fig:differenze}, which shows that the separation among curves with different $\Delta Y/\Delta Z$ becomes progressively smaller and smaller as the [Fe/H] decreases. At low enough values of [Fe/H], this separation is simply too small 
with respect to the spread of the observed point, due to the current uncertainties in both magnitudes and [Fe/H], which are of the order of 0.1 mag and 0.1 dex respectively, to allow a correct determination of the stars chemical composition. This uncertainty is reflected in the large error bars of the lower metallicity bins of Fig. \ref{fig:binned}.

We are not claiming that any non linear law for the helium-to-metals enrichment of the ISM  must be ruled out; we are just stating that better data are needed to confirm this possibility by making use of the fine structure of the low main sequence. The GAIA mission will provide us
 with better parallaxes, which will yield  better absolute magnitude determinations.
 Advances in 3D and NLTE stellar atmospheres modeling will also produce better constrained spectroscopic abundances determinations in the future.

\section{The Hyades test}
\label{sec:Hyatest}

As previously discussed, the $\Delta Y/\Delta Z$ provided by the recovery procedure must be corrected for the age-bias, since the above described technique uses the theoretical ZAMS loci as reference mile stones in the CMD. Hence, any unaccounted displacements of the 
observed stars from the ZAMS  due to evolutionary effects translates directly in an underestimate of the inferred $\Delta Y/\Delta Z$. Clearly, the younger the stellar sample analyzed, the smaller such unaccounted displacements and the smaller the related age-bias. 

To safely neglect any possible bias due to evolution, we also applied our recovery method to a very young stellar sample: the stars of the lowest part ($M_\mathrm{V} \ge 5.2$) of the Hyades MS. 
For this well studied cluster an age of about $500 \div 600$ Myr is generally estimated \citep[see e.g.][]{perryman98, castellani01,castellani02}. Thus, the selected low luminosity stars, which correspond to masses lower than about $0.9 M_{\sun}$, are really on, or at least very close to, their ZAMS position.
Moreover, the data available for this cluster are even more precise than those of our data set for field stars in the solar neighborhood; using the kinematically corrected parallaxes data from \citet{Madsen02} the typical errors are $\sigma (M_\mathrm{V}) \simeq 0.05 \,\mbox{mag}$ and $\sigma (B -V) \simeq0.02 \,\mbox{mag}$.
The metallicity value we use for the Hyades is taken from \citet{perryman98}: [Fe/H] = $0.14 \pm 0.05 \,\mbox{dex}$.

By running our Monte Carlo method, the value found for the enrichment ratio is $(\Delta Y/ \Delta Z)_{Hyad} = 4.75 \pm 0.35$, in perfect agreement with what we found using local MS stars and taking into account the evolutionary bias.

To further check the consistency of this final result, we calculated isochrones using the closest value of the enrichment ratio in our grid, i.e. $\Delta Y/ \Delta Z =5$, and for the extreme values of \cite{perryman98} [Fe/H] interval, i.e. 0.09 and 0.19 dex.
Models are calculated for our solar calibrated value of the mixing length parameter and transformed to the observational plane using PHOENIX model atmosphere.
Average extinction towards the Hyades is negligible and this cluster is old enough not to present any intra-cluster material, so also differential extinction can be ignored; hence we assume $(B - V)_{0} = (B - V)_{obs}$.

\begin{figure}
   \centering
\resizebox{\hsize}{!}{\includegraphics{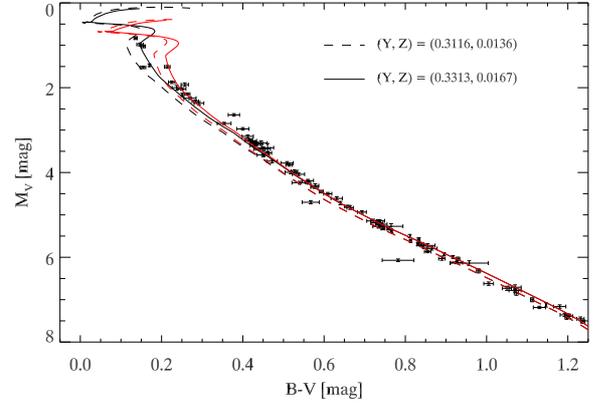}}
   \caption{The CMD for the Hyades Main Sequence. Isochrones for two values of the age are overplotted: 500 Myr (black lines) and 600 Myr (red lines). Dashed and solid lines correspond to two different values of the chemical composition. $Y$ and $Z$ values are obtained from Equations (\ref{eq:dydzlin}) and (\ref{eq:zeta}) using [Fe/H] = 0.09 dex (dashed lines) and [Fe/H] = 0.19 dex (solid lines).}
    \label{fig:IADIMS}
 \end{figure}

Figure \ref{fig:IADIMS} shows the CMD of the Hyades with overimposed our isochrones.
The good agreement between our models with the chemical composition derived by the recovery method and the Hyades data is quite encouraging and make us confident about the adopted procedure.
Even more important is that the fit to the Hyades MS is good also at higher luminosities.
Stars with $M \gtrsim 1.2 M_{\sun}$ (corresponding to a magnitude $M_\mathrm{V} \lesssim 3.6$, the exact value depending on chemical composition and age) have essentially radiative envelopes, meaning that their predicted temperatures are not affected by the choice on the mixing length parameter, $\alpha$.

\section{Comparison with independent methods}
\label{sec:othermeth}

Many independent approaches have been followed in the past to determine 
the helium-to-metals enrichment ratio. 
A detailed comparison between the value of helium-to-metal enrichment
 ratio $\Delta Y/ \Delta Z$ obtained by means of different techniques 
is beyond the aim of the paper, since it would require a careful
 discussion of the different uncertainty sources, both systematics and random,
 affecting the various techniques. Furthermore, the primordial helium 
abundance $Y_p$ adopted by different authors and/or in different periods
 can be very different, affecting the inferred $\Delta Y/ \Delta Z$ value. 
On the other hand, we think that a brief description of the results provided
 by different approaches might be of interest.

One of the most fruitful consists in observing HII regions in the Milky Way
and in other galaxies. The pioneering papers by \citet{peimbert74,peimbert76} 
and by \citet{lequeux79}, devoted respectively to the HII regions in the Magellanic Clouds
and in irregular and blue compact galaxies, derived a value of 
the helium to metals enrichment ratio of $\Delta Y / \Delta Z\approx 3$. 
From observations of extragalactic HII regions, \citet{pagel92} found that 
$\Delta Y / \Delta Z \geq 3$, with a preferred value close to 4, while
\citet{peimbert00} obtained $\Delta Y / \Delta O = 3.5 \pm 0.9 $ from NGC346, the 
brightest HII region in SMC. 
A detailed study of a very large sample of spectroscopic observations of 
HII regions in blue compact galaxies performed by \citet{izotov04} led to 
 $\Delta Y / \Delta Z = 2.8 \pm 0.5 $. However, a reanalysis of the same data
 by \citet{fukugita06} provides $\Delta Y /\Delta Z= 4-5$, values that becomes
 as low as 1.1$\pm$ 1 if stellar absorption is taken into account. 

From observations of galactic HII regions, such as the Orion nebula, M8 and M17, various
authors found $\Delta Y /\Delta Z> 2$  \citep[see also][]{esteban99,deharveng00}. 
As an example \citet{peimbert00} found $\Delta Y / \Delta Z = 2.1 \pm 0.6 $
adopting a temperature fluctuation parameter $t^2=0.037$ (recommended) 
and $3.8 \pm 1.1 $ adopting $t^2=0.000$.  
More recently, \citet{balser06} derived $\Delta Y / \Delta Z = 1.41 \pm 0.62$
 from the analysis of the sole M17 and S206, 
whereas taking into account also the HII regions belonging to the Magellanic
Clouds and metal-poor galaxies, she found $\Delta Y /\Delta Z= 1.6$. The
last author warns also on the systematic underestimation of the measured helium abundance 
caused by a clumping in HII regions \citep[see also][]{mathis05}.   
Finally, a new estimate of the chemical composition of M17 by \citet{carigi08}
yields $\Delta Y / \Delta Z = 1.97 \pm 0.41$ and $4.00 \pm 0.75$ adopting
$t^2=0.036$ and 0.000, respectively. 

A useful technique to empirically constrain the stellar helium to heavier
elements enrichment ratio is the analysis of the chemical composition of planetary
nebulae (PNe). These objects present the advantage of being
numerous in a large range of metallicity and of allowing precise estimates
 of helium abundance.
On the other hand, it is a well established result of stellar evolution that
the chemical composition of low and intermediate mass stars, i.e. the PNe's
progenitors, is strongly modified by the dredge-up episodes, which enrich 
the envelope with inner material previously processed by nuclear reactions,
that is mainly with fresh helium and, in the case of III dredge-up in thermally
pulsating AGB stars, also with heavier elements, mainly carbon.   
Thus, the chemical composition of PNe is not representative of the
protostellar cloud and in order to derive the original interstellar abundances
 a correction must be applied which takes into account the evolution of the
 surface abundances of the progenitor star. 
In the early attempts to determine the $\Delta Y /\Delta Z$ ratio using the
PNe in the late '70s, such a correction was neglected and values in the range 
2.2-3.6 were obtained \citep[see e.g.][]{dodorico76,peimbert80}.  
In our knowledge, \citet{chiappini94} were the first to take into 
account the correction due to the evolution of the progenitor star and
they obtained  $4~\leq\Delta Y /\Delta Z\leq~6.3$.
More recently, \citet{maciel01} found $2.8~\leq\Delta Y / \Delta Z\leq~3.6$
adopting $Y_\mathrm{P}=0.23$ and $2.0~\leq\Delta Y / \Delta Z\leq~2.8$ with $Y_\mathrm{P}=0.24$.

An important independent constraint on the stellar helium to heavier
elements enrichment ratio is provided by the Sun, for which very detailed and
accurate data are available. In order to derive the $\Delta Y /\Delta Z$
ratio from the Sun, the precise spectroscopic estimates of the solar
photospheric chemical composition are not enough because they significantly
 differ from the original ones due to diffusion and gravitational settling. 
Thus, one has to rely on a standard solar model (SSM), that is, a stellar
model of 1 $M_{\sun}$ which at the age of the Sun (i.e. $\approx$4.56 Gyr) fits the 
solar observables. As a result of such a procedure, the initial metallicity
$Z_{\sun}$ and helium $Y_{\sun}$ abundance are inferred. As largely debated
in recent years \citep[see e.g][and references therein]{bahcall05b,basu08,christensen09,serenelli09},
 the very good agreement between the SSM and the helioseismological
 constraints, mainly the sound speed profile, the extension
of the convective envelope and the surface helium abundance,
 achieved at the end of the last century has been compromised by
 the new determinations of the metal abundances based on 3D photospheric
 models by \citet{aspl05}, and only slightly
 alleviated by the very recent release by \citet{aspl09}.   
With these caveats, the initial helium and metal abundance of the Sun provided
by the current SSMs allow to constrain the $\Delta Y /\Delta Z$, once a
primordial helium abundance $Y_\mathrm{P}$ is chosen. Table \ref{tab:solabb} shows the results of
SSMs computed by different groups by assuming different solar heavy elements
mixtures. As can be easily seen, different authors provide values in good mutual
agreement, values that translate in  $\Delta Y /\Delta Z<~1$, if $Y_\mathrm{P}=0.248$
by \citet{izotov07} is used.  


\begin{table}
\begin{minipage}[t]{\columnwidth}
\caption{Initial helium $Y_{ini}$ and metal $Z_{ini}$ abundances provided by 
recent standard solar models by different authors (see footnotes). Models are computed adopting the heavy element mixture by
\citet{aspl05}.}
\label{tab:solabb}
\centering
\renewcommand{\footnoterule}{}  
\begin{tabular}{|c|ccc|}
\hline
\hline
 Models & $Y_\mathrm{ini}$ & $Z_{ini}$ & $\Delta Y/ \Delta Z$ \\
\hline
 Pisa\footnote{\citet{tognelli10}}  & 0.2532 & 0.0137  &  0.380 \\
 B2005\footnote{\citet{bahcall05}} & 0.2614 & 0.0140  &  0.957 \\
 G2006\footnote{\citet{guzik05}} & 0.2570 & 0.0135  &  0.666 \\
 S2009\footnote{\citet{serenelli09}} & 0.2593 & 0.0139  &  0.813 \\
 S2009b\footnote{\citet{serenelli09}, mixture by \citet{aspl09}} & 0.2617 & 0.0149  &  0.919 \\
\hline

\end{tabular}
\end{minipage}
\end{table}

In principle the $\Delta Y / \Delta Z$ ratio can also be theoretically 
predicted by models of galactic chemical evolution.
These predictions rely on several assumption on the initial mass function, star formation rate, star formation efficiency, stellar yields from intermediate and massive stars, ISM mixing; so even though they are quite powerful, due to the uncertainties on each of the aforementioned quantities they are still quite uncertain in predicting the expected helium to metals enrichment ratio and provide values
  which span a range of $\Delta Y / \Delta Z$ between about 1 and 4
 \citep{serrano81,timmes95,tosi96,chiappini97,fields98,carigi00,chiappini02,romano05,carigi08}. 
Among the parameters that mostly affect the outcomes of chemical evolution models we mention some of the most important, such as the SN Ia rate, which strongly affects the iron enrichment, the star formation history and the initial mass function, since stars of different mass, which evolve on different time scales, contribute differently to the production of $Y$ and $Z$; the maximum stellar mass that enriches the interstellar medium is an other important parameter. 
Moreover, as far as the chemical enrichment of the interstellar medium is concerned, stellar models, which provide the yields to
chemical models, are still significantly affected by poorly understood  physical processes, such as convection, which affects the stellar  yields
 through the efficiency of dredge-up phenomena and hot-bottom burning, mass-loss, which directly influence the enrichment, and rotation,
 which has a strong impact on the helium, carbon, nitrogen and oxygen yields \citep{meynet02}.
 As a consequence, stellar yields provided by different authors still show not negligible discrepancies.
 Thus, the results of chemical evolution models depend on the adopted set of
 stellar yields \citep{fields98,carigi00, chiappini03}.
 For example, it has long been known that the $\Delta Y / \Delta Z$ ratio grows by increasing the mass-loss rate of both 
intermediate and massive stars \citep[see e.g.][]{iben78,renzini81,peimbert86}.   

\citet{chiappini03} predicted a value of the enrichment ratio $\Delta Y /
\Delta Z \sim~2.4$, if the stellar models by \citet{meynet02}, which take into
account rotation, are adopted in their chemical evolution model,
 whereas a value of 1.5 is obtained, if the stellar computations
 by \citet{vandenhoek97}, for the low and intermediate mass stars, and
 \citet{woosley95}, for the massive ones, are used.    
The recent version of the chemical evolution model by \citet{carigi08}      
provides $\Delta Y / \Delta Z =$1.70 and 1.62, if high or low wind yields are
 adopted, respectively.

\section{Conclusions}
\label{sec:concl}

The principal aim of this work was to test the reliability of the
determination of $\Delta Y/ \Delta Z$ by the comparison between low-MS stars and
 theoretical ZAMS models. A very fine grid of stellar models has been 
computed for many values of $\Delta Y/ \Delta Z$, [Fe/H] and masses
 adopting two different mixing-length parameters $\alpha$, namely 1.97 (our solar calibration) 
and 2.4, and two different sets of atmosphere models to transform 
luminosities and effective temperatures into magnitudes and color indices, the PHOENIX GAIA v2.6.1  
models \citep{brott} and the ATLAS9 ones \citep{castelli03}, respectively. 

A detailed analysis of the capabilities of the method and of the main uncertainty sources 
affecting the derived results has been performed by means of many numerical 
experiments on synthetic data set produced under controlled conditions and with 
precisely known properties.

One of the main findings of the paper is that the inferred value of
 $\Delta Y/ \Delta Z$ is quite sensitive to the age of the stellar sample,
 even in the case in which only very faint (i.e. $M_V > 6$ mag) MS stars are selected. 
By means of numerical experiment we showed that the lack of an age estimate of
 low mass field stars leads to an underestimate of the inferred $\Delta Y/ \Delta Z$ of about $2$ units. 
As a consequence, the face value of the helium-to-metals enrichment ratio
provided by the recovery procedure applied to the solar neighborhood stellar sample 
must be corrected for this age-bias. 

Adopting our reference set of models (i.e. those with our ``solar'' calibrated $\alpha$)
transformed into the observational plane using the PHOENIX GAIA v2.6.1 model atmospheres, we 
found $\Delta Y/\Delta Z = 5.3 \pm 1.4$. 

Such a result has been checked against an independent and very accurate data set, that is, the 
low-MS of the Hyades cluster. This sample has the additional advantage to be unaffected 
by the age-bias, since its very young age (i.e. $500 \div 600$ Myr) guarantees that the low-mass 
($<0.9 M_{\sun}$) MS stars which constitutes the sample are essentially unevolved.  
The recovery method provided again $\Delta Y/\Delta Z = 4.75 \pm 0.35$, in
perfect agreement with the result for
field stars.

To further check the consistency of this final result, we calculated isochrones with
 $\Delta Y/ \Delta Z = 5$ (the closest in our grid of models) and the measured [Fe/H] of Hyades. The good agreement 
between these isochrones and the Hyades MS, not only in the faint part belonging to the sample 
but also at higher luminosities, is a further proof of the internal consistency of the 
recovery procedure. 

The effect of a change in the assumed efficiency of the superadiabatic convection (i.e. the $\alpha$ 
parameter) in the stellar models used to build the ZAMS and in the adopted atmosphere models 
used to transform luminosities and effectives temperatures into magnitudes and color indices has 
been discussed, too. More in detail, the recovery method yield a nominal value
of $\Delta Y/\Delta Z=  1.59 \pm 1.01$ when adopting the set of theoretical models computed
with $\alpha=~2.4$ and transformed into the observational plane by means of
PHOENIX atmosphere models and $\Delta Y/\Delta Z= 0.5$ when adopting our standard set
of models with $\alpha=~1.97$ but the ATLAS9 atmosphere models.
These values become about 3.6 and 2.5, respectively, once corrected for the
age bias. 

Our data have [Fe/H] determinations coming from two different sources, i.e. \cite{Nordstrom04} and \cite{Taylor05} catalogs.
The 4 stars we have in common from the two catalogs show a disagreement in the [Fe/H] scale between the two catalogs. This disagreement was already identified in \cite{Taylor05} and the author provides a Table to account for it and put the two catalogs on the same [Fe/H] scale. Nevertheless, even after the re-zeroing procedure suggested by \cite{Taylor05}, we still find some disagreement in the final results on $\Delta Y/ \Delta Z$ when the two catalogs are considered separately. Probably an even more accurate study of the zero points of metallicity determinations is needed.

\begin{acknowledgements}
We are very grateful to Emanuele Tognelli for his contribution to the updating
of the code and to Steven N. Shore for the many pleasant discussions on the topic. 
 PGPM thanks Wolfgang Brandner for the kind hospitality at
Max-Planck-Institute for Astronomy, Heidelberg, where a part of the paper has been written. 
This work has been supported by PRIN-MIUR 2007 ({\em Multiple stellar
  populations in globular clusters: census, characterizations and origin}, PI G. Piotto) 
\end{acknowledgements}

\bibliographystyle{aa}

\begin{thebibliography}{}
\bibitem[{Asplund et al.(2005)}]{aspl05} Asplund, M., Grevesse, N., \& Sauval, A.J. 2005, in ``Cosmic Abundances as Records of Stellar Evolution and Nucleosynthesis'', eds. F.N. Bash, \& T.J. Barnes, ASP Conf. Series, 336, 25
\bibitem[{Asplund et al.(2009)}]{aspl09} Asplund, M., Grevesse, N., Sauval,
  A. J., Scott, P. 2009, ARA\&A, 47, 481	
\bibitem[{Bahcall et al. (2005b)}]{bahcall05b} Bahcall, J. N., Basu, Pinsonneault, M., S.Serenelli, A. M. 2005b, ApJ, 618, 1049  
\bibitem[{Bahcall, Serenelli \& Basu
    (2005)}]{bahcall05} Bahcall, J. N., Serenelli, A. M., \& Basu, S. 2005, ApJ, 621, L85
\bibitem[{Balser (2006)}]{balser06} Balser, D. S. 2006, AJ, 132, 2326 
\bibitem[{Basu \& Antia (2008)}]{basu08} Basu, S., \& Antia, H. M. 2008, Phys. Rep., 457, 217
\bibitem[{B\"{o}hm-Vitense(1958)}]{bohm58} B\"{o}hm-Vitense, E. 1958, Zs.f.Ap., 46, 108
\bibitem[{Bessel, Castelli \& Plez (1999)}]{bessel99} Bessell, M.~S., Castelli, F. \& Plez, B. 1999, A\&A, 333, 231
\bibitem[{Brott et al.(2005)}]{brott} Brott, I., \& Hauschildt, P. H. 2005, ESA Special Publication, 576, 565
\bibitem[{Carigi (2000)}]{carigi00} Carigi, L. 2000, RMxAA, 36, 171
\bibitem[{Carigi \& Peimbert (2008)}]{carigi08} Carigi, L., \& Peimbert, M. 2008, RMxAA, 44, 341
\bibitem[{Casagrande et al.(2007)}]{casagrande07} Casagrande, L., Flynn, C., Portinari, L., Girardi, L., Jimenez, R. 2007, MNRAS, 382, 1516
\bibitem[{Castellani, Degl'Innocenti \& Marconi (1999)}]{castellani99} Castellani, V., Degl'Innocenti, S., \& Marconi, M. 1999, A\&A, 349, 834 
\bibitem[{Castellani, Degl'Innocenti \& Prada Moroni (2001)}]{castellani01} Castellani, V., Degl'Innocenti, S., \& Prada Moroni, P. G. 2001, MNRAS, 320, 66
\bibitem[{Castellani et al. (2002)}]{castellani02} Castellani, V., Degl'Innocenti, S., Prada Moroni, P. G., Tordiglione, V. 2002, MNRAS, 334, 193
\bibitem[{Castelli \& Kurucz (2003)}]{castelli03} Castelli, F. \& Kurucz, R.~L., IAU Symposium, 210, 20
\bibitem[{Chiappini \& Maciel (1994)}]{chiappini94} Chiappini, C., \& Maciel, W. J. 1994, A\&A, 288, 921
\bibitem[{Chiappini, Matteucci \& Gratton (1997)}]{chiappini97} Chiappini, C., Matteucci, F., \& Gratton, R. 1997, ApJ, 477, 765
\bibitem[{Chiappini, Matteucci \& Meynet (2003)}]{chiappini03} Chiappini, C., Matteucci, F., Meynet, G. 2003, A\&A, 410, 257
\bibitem[{Chiappini, Renda \& Matteucci (2002)}]{chiappini02} Chiappini, C.; Renda, A.; Matteucci, F. 2002, A\&A, 395, 789
\bibitem[{Chieffi \& Straniero(1989)}]{chieffi} Chieffi, A., \& Straniero, O. 1989, ApJS, 71, 47
\bibitem[{Christensen-Dalsgaard et al. (2009)}]{christensen09} Christensen-Dalsgaard, J., di Mauro, M. P., Houdek, G., Pijpers, F. 2009, A\&A, 494, 205
\bibitem[{Churchwell, Mezger \& Huchtmeier (1974)}]{churchwell74} Churchwell, E., Mezger, P. G., \& Huchtmeier, W. 1974, A\&A, 32, 283
\bibitem[{Deharveng et al. (2000)}]{deharveng00} Deharveng, L., Pena, M., Caplan, J., Costero, R. 2000, MNRAS, 311, 329
\bibitem[{Degl'Innocenti, Prada Moroni \& Ricci (2006)}]{deg06} Degl'Innocenti, S., Prada Moroni, P.G., \& Ricci, B. 2006, Ap\&SS, 305, 67 
\bibitem[{Degl'Innocenti et al.(2008)}]{deg08} Degl'Innocenti, S., Prada Moroni, P.G., Marconi, M., \& Ruoppo, A. 2008, Ap\&SS, 316, 25
\bibitem[{Dodorico, Peimbert \& Sabbadin (1976)}]{dodorico76} Dodorico, S., Peimbert, M., \& Sabbadin, F. 1976, A\&A, 47, 341
\bibitem[{Dunkley et al.(2009)}]{dunkley09} Dunkley, J., et al. 2009, ApJS, 180, 306
\bibitem[{ESA (1997)}]{HIP} The Hipparcos and Tycho Catalogues (ESA 1997)
\bibitem[{Esteban at al.(1999)}]{esteban99} Esteban, C., Peimbert, M., Torres-Peimbert, S., Garcia-Rojas, J. 1999, RMxAA, 35, 65
\bibitem[{Faulkner (1967)}]{faulkner67} Faulkner, J., 1967, ApJ, 147, 617 
\bibitem[{Fields \& Olive (1998)}]{fields98} Fields, B. D., \& Olive, K. A. 1998, ApJ, 506, 177
\bibitem[{Ferguson et al.(2005)}]{ferg} Ferguson, J.W., Alexander, D.R., Allard, F., et al. 2005, ApJ, 623, 585
\bibitem[{Fernandes, Lebreton \& Baglin (1996)}]{fernandes96} Fernandes, J., Lebreton, Y., Baglin, A. 1996, A\&A, 311, 127
\bibitem[{Fukugita \& Kawasaki (2006)}]{fukugita06} Fukugita, M., \& Kawasaki, M. 2006, ApJ, 646, 691
\bibitem[{Girardi et al.(2002)}]{girardi02} Girardi, L., Bertelli, G., Bressan, A. et al. 2002, A\&a, 391, 195
\bibitem[{Grevesse \& Noels (1993)}]{grev93} Grevesse, N., \& Noels, A.\ 1993, Origin and Evolution of the Elements, 15 
\bibitem[{Guzik, Watson \& Cox (2005)}]{guzik05} Guzik, J. A., Watson, L. S.,
  \& Cox, A. N. 2005, ApJ, 627, 1049
\bibitem[{Holmberg et al.(2007)}]{Holmberg07} Holmberg, J., Nordstr{\"o}m, B. \& Andersen, J. 2007, A\&A, 475, 519
\bibitem[{Iben \& Truran (1978)}]{iben78} Iben, I., Jr., \& Truran, J. W. 1978, ApJ, 220, 980
\bibitem[{Iglesias \& Rogers(1996)}]{ig96} Iglesias, C., \& Rogers, F.J. 1996,  ApJ, 464, 943
\bibitem[{Izotov \& Thuan (2004)}]{izotov04} Izotov, Y. I., \& Thuan, T. X. 2004, ApJ, 602, 200
\bibitem[{Izotov, Thuan \& Lipovetsky (1994)}]{izotov94} Izotov, Y. I., Thuan, T. X., \&  Lipovetsky V. A. 1994, ApJ, 435, 647
\bibitem[{Izotov, Thuan \& Lipovetsky (1997)}]{izotov97} Izotov, Y. I., Thuan, T. X., \&  Lipovetsky V. A. 1997, ApJS, 108, 1 
\bibitem[{Izotov, Thuan \& Stasi (2007)}]{izotov07} Izotov, Y. I., Thuan, T. X., \& Stasi{\'n}ska, G. 2007,  ApJ, 662, 15
\bibitem[{Jimenez et al. (2003)}]{jimenez03} Jimenez, R., Flynn,  C., MacDonald,  J., Gibson,  B. K. 2003, Science, 299, 1552  
\bibitem[{Krishna Swamy (1966)}]{krishna66}  Krishna Swamy, K. S. 1966, ApJ, 145, 174
\bibitem[{Kroupa(2001)}]{Kroupa01} Kroupa, P. 2001, MNRAS, 322, 231 
\bibitem[{Kunth (1986)}]{kunth86} Kunth, D. 1986, PASP, 98, 984
\bibitem[{Kunth \& Sargent (1983)}]{kunth83} Kunth, D., \& Sargent, W. L. W. 1983, ApJ, 273, 81
\bibitem[{Lequeux et al. (1979)}]{lequeux79} Lequeux, J., Peimbert, M., Rayo, J. F., Serrano, A., Torres-Peimbert, S. 1979, A\&A, 80, 155
\bibitem[{Maciel (2001)}]{maciel01} Maciel, W. J. 2001, Ap\&SS, 277, 545
\bibitem[{Madsen, Dravins \& Lindegren (2002)}]{Madsen02} Madsen, S., Dravins, D. \& Lindegren, L. 2002, A\&A, 381, 446
\bibitem[{Mathews, Boyd \& Fuller (1993)}]{mathews93} Mathews, G. J., Boyd, R. N., \& Fuller, G. M. 1993, ApJ, 403, 65
\bibitem[{Mathis \& Wood (2005)}]{mathis05} Mathis, J. S., \& Wood, K. 2005, MNRAS, 360, 227
\bibitem[{Meynet \& Maeder (2002)}]{meynet02} Meynet, G., \& Maeder, A. 2002, A\&A, 390, 561
\bibitem[{Montalban et al. (2004)}]{montalban04} Montalban, J., D'Antona, F., Kupka, F., Heiter, U. 2004, A\&A, 416, 1081
\bibitem[{Nordstr{\"o}m et al.(2004)}]{Nordstrom04} Nordtr{\"o}m B., Mayor, M., Andersen, J. et al 2004, A\&A, 418, 989
\bibitem[{Olive \& Steigman (1995)}]{olive95} Olive, K. A., \& Steigman, G. 1995, ApJS, 97, 49
\bibitem[{Olive, Steigman \& Skillman (1997)}]{olive97} Olive, K. A., Steigman, G., \& Skillman, E. D. 1997, ApJ, 483, 788
\bibitem[{Pagel \& Portinari (1998)}]{pagel98} Pagel, B. E. J., \& Portinari, L. 1998, MNRAS, 298, 747
\bibitem[{Pagel, Terlevich \& Melnick (1986)}]{pagel86} Pagel, B. E. J., Terlevich, R. J., Melnick, J. 1986, PASP, 98, 1005
\bibitem[{Pagel et al. (1992)}]{pagel92} Pagel, B. E. J., Simonson, E. A., Terlevich, R. J., Edmunds, M. G. 1992, MNRAS, 255, 325
\bibitem[{Peebles (1966)}]{peebles66} Peebles, P. J. E. 1966, ApJ, 146, 542
\bibitem[{Peimbert (1986)}]{peimbert86} Peimbert, M. 1986, PASP, 98, 1057
\bibitem[{Peimbert, Carigi \& Peimbert (2001)}]{peimbert01} Peimbert, M., Carigi, L., \& Peimbert, A. 2001, ApSSS, 277, 147	
\bibitem[{Peimbert \& Serrano (1980)}]{peimbert80} Peimbert, M., \& Serrano, A. 1980, RMxAA, 5, 9
\bibitem[{Peimbert \& Torres-Peimbert (1974)}]{peimbert74} Peimbert, M., \& Torres-Peimbert, S. 1974, ApJ, 193, 327
\bibitem[{Peimbert \& Torres-Peimbert (1976)}]{peimbert76} Peimbert, M., \& Torres-Peimbert, S. 1976, ApJ, 203, 581
\bibitem[{Peimbert, Peimbert \& Luridiana (2002)}]{peimbert02} Peimbert, A., Peimbert, M., \& Luridiana, V. 2002, ApJ, 565, 668
\bibitem[{Peimbert et al. (2000)}]{peimbert00} Peimbert, M., Peimbert, A., \& Ruiz, M. T. 2000, ApJ, 541, 688
\bibitem[{Peimbert et al. (2007)}]{peimbert07} Peimbert, M., Luridiana, V., Peimbert, A.,  \& Carigi, L. 2007, Astr. Soc. of the Pac. Conf. Series, 374, 81
\bibitem[{Perrin et al. (1977)}]{perrin77} Perrin, M.-N., de Strobel, G. C., Cayrel, R., Hejlesen, P. M. 1977, A\&A, 54, 779 
\bibitem[{Perryman et al. (1998)}]{perryman98} Perryman, M.~A.~C., Brown, A.~G.~A., Lebreton, Y. et al. 1998, A\&A, 331, 81
\bibitem[{Potekhin (1999)}]{pot} Potekhin, A.Y. 1999, A\&A, 351, 787
\bibitem[{Renzini (1994)}]{renzini94} Renzini, A. 1994, A\&A, 285L, 5
\bibitem[{Renzini \& Voli (1981)}]{renzini81} Renzini, A., \& Voli, M. 1981, A\&A, 94, 175
\bibitem[{Rogers et al.(1996)}]{rog96} Rogers, F.J., Swenson, F.J., \& Iglesias, C.A. 1996, ApJ, 456, 902
\bibitem[{Romano et al.(2005)}]{romano05} Romano, D., Chiappini, C., Matteucci, F., Tosi, M. 2005, A\&A, 430, 491
\bibitem[{Salpeter (1955)}]{Salpeter55} Salpeter, E.~E.\ 1955, ApJ, 121, 161 
\bibitem[{Serenelli et al.(2009)}]{serenelli09} Serenelli, A. M., Basu, S., Ferguson, J. W., Asplund, M. 2009, ApJ, 705, L123
\bibitem[{Serrano \& Peimbert (1981)}]{serrano81} Serrano, A., \& Peimbert, M. 1981, RMxAA, 5, 109
\bibitem[{Shternin \& Yakovlev(2006)}]{shtern} Shternin, P.S. \& Yakovlev, D.G. 2006, PhRvD, 74(4), 3004
\bibitem[{Spergel et al. (2007)}]{spergel07} Spergel, D. N., et al. 2007, ApJS, 170, 377
\bibitem[{Taylor (2005)}]{Taylor05} Taylor, B.~J., ApJS, 161, 444
\bibitem[{Timmes, Woosley \& Weaver (1995)}]{timmes95} Timmes, F. X., Woosley, S. E., \& Weaver, T. A. 1995, ApJS, 98, 617
\bibitem[{Tognelli, Prada Moroni \& Degl'Innocenti (2010)}]{tognelli10} Tognelli, E., Prada Moroni, P. G., \& Degl'Innocenti, S. 2010, A\&A, in press
\bibitem[{Tosi (1996)}]{tosi96} Tosi, M. 1996, ASP Conf. Ser., 98, 299
\bibitem[{Valle et al. (2009)}]{valle09} Valle, G., Marconi, M.,
  Degl'Innocenti, S., Prada Moroni, P. G. 2009, A\&A, in press
\bibitem[{van den Hoek \& Groenewegen (1997)}]{vandenhoek97} van den Hoek, L. B., \& Groenewegen, M. A. T. 1997, A\&AS, 123, 305
\bibitem[{Woosley \& Weaver (1995)}]{woosley95} Woosley, S. E., \& Weaver, T. A. 1995, ApJS, 101, 181
\end{thebibliography}

\end{document}